\documentclass[aps,prd,superscriptaddress,nofootinbib,amsmath,amsfonts,preprintnumbers,groupedaddress,showpacs,10pt,english]{revtex4-1}
\usepackage{amsmath}
\usepackage{amssymb}
\usepackage{babel}
\usepackage{wrapfig}
\usepackage{cancel}

\usepackage{relsize,exscale}
\makeatletter

\usepackage{array,multirow,graphicx}
\usepackage{dcolumn}
\usepackage{newlfont}
\usepackage{bm}
\usepackage[colorlinks,citecolor=blue,urlcolor=blue,linkcolor=blue]{hyperref}
\usepackage[figtopcap]{subfigure}
\usepackage{color}

\newcommand\be{\begin{equation}}
\newcommand\ba{\begin{eqnarray}}
\newcommand\ee{\end{equation}}
\newcommand\ea{\end{eqnarray}}

\begin{document}
\tolerance=5000

\title{  New anisotropic star solutions in   mimetic gravity}
\author{G.~G.~L.~Nashed$^{1,2}$}\email{nashed@bue.edu.eg}
\author{and Emmanuel N. Saridakis$^{3,4,5}$}\email{msaridak@noa.gr}

\affiliation {$^1$ Centre for Theoretical Physics, The British University, P.O. Box
43, El Sherouk City, Cairo 11837, Egypt \\
$^2$Center for Space Research, North-West University, Mahikeng 2745, South Africa\\
$^3$National Observatory of Athens, Lofos Nymfon, 11852 Athens,
Greece\\
$^4$ CAS Key Laboratory for Researches in Galaxies and Cosmology,
Department of Astronomy, University of Science and Technology of China, Hefei,
Anhui 230026, P.R. China\\
$^5$Departamento de Matem\'{a}ticas, Universidad Cat\'{o}lica del
Norte, Avda. Angamos 0610, Casilla 1280 Antofagasta, Chile}



\begin{abstract}
We extract new classes of anisotropic solutions in the framework of   mimetic
gravity, by applying the Tolman-Finch-Skea metric  and a specific anisotropy
not directly depending on  it, and by matching smoothly the interior
anisotropic solution to the Schwarzschild exterior one. Then, in order to
provide a
transparent picture  we use    the data from the 4U
1608-52 pulsar. We study the profile of the energy density,
as well as the  radial and tangential pressures, and we show that they are all
positive and decrease towards the center of the star. Furthermore, we
investigate the anisotropy parameter and the anisotropic force, that are both
increasing functions of the radius, which implies that the latter is
repulsive. Additionally, by examining the radial and tangential
equation-of-state parameters, we show that they are   monotonically
increasing, not corresponding to exotic matter.
Concerning the  metric potentials we find that they have no singularity, either
at the center of the star or at the boundary. Furthermore, we verify that all
energy conditions are satisfied, we show that the   radial and tangential
sound speed squares   are positive and sub-luminal, and we
find that the surface red-shift satisfies the theoretical requirement. Finally,
in order to investigate the stability    we
apply the  Tolman-Oppenheimer-Volkoff   equation, we perform the
adiabatic index analysis, and we examine the static case,  showing that in all
cases the
star is stable.
\end{abstract}

\keywords{Mimetic gravity, anisotropic solutions, compact stars, TOV
equation, adiabatic index.}
\maketitle

\section{Introduction}\label{S1}

Astrophysical compact objects, such as neutron stars  and black holes, can
serve as a crucial  laboratory to investigate gravitational fields in the
strong-field regime, and thus test General Relativity and its possible
extensions
\cite{Damour:1996ke,Will:2014kxa,Berti:2015itd,
LIGOScientific:2016lio,  Berti:2018cxi}.
Such extensions usually arise through the consideration of higher-order terms
in the Einstein-Hilbert Lagrangian, such as in $fR)$   gravity
\cite{DeFelice:2010aj}, in  Gauss-Bonnet and $f(G)$ gravity
 \cite{Antoniadis:1993jc,Nojiri:2005jg}, in Weyl gravity
\cite{Mannheim:1988dj}, in   Lovelock and  f$($Lovelock$)$  gravity
\cite{Lovelock:1971yv,Deruelle:1989fj}, in scalar-tensor theories
\cite{Fierz:1956zz,Jordan:1959eg,Brans:1961sx,Damour:1992we}
etc (for a review see \cite{CANTATA:2021ktz}). Additionally, one may construct
different classes of modifications by modifying   the equivalent, torsional
formulation of gravity, resulting in  $f(T)$ gravity
\cite{Bengochea:2008gz,Cai:2015emx}, in $f(T,T_G)$ gravity
\cite{Kofinas:2014owa}, in
scalar-torsion theories \cite{Geng:2011aj} etc.
Hence, in the literature one may find many studies of spherically symmetric
solutions in the framework of modified gravity
\cite{Garfinkle:1990qj,Kanti:1995vq,Cai:2001dz,Kanti:2004nr,Emparan:2008eg,Nojiri:2010wj,
Kiritsis:2009rx, Gonzalez:2011dr,Pani:2011mg,Pani:2011xm,Capozziello:2012zj,Nojiri:2017ncd,
Anabalon:2013oea,Garattini:2014rwa,Cai:2012db, Nashed:2013bfa,Astashenok:2015qzw,
Cisterna:2014nua,Paliathanasis:2014iva,Lu:2015cqa,Moraes:2015uxq,Astashenok:2015haa,
Babichev:2015rva,
Erices:2017izj,Doneva:2017jop,Doneva:2017bvd,Jasim:2018cos,Roupas:2020nua,
Karakasis:2021rpn, Nashed:2018efg,
Ren:2021uqb,Mota:2022zbq,Nashed:2018cth,Kiorpelidi:2022kuo,Chatzifotis:2022mob,Zhao:2022gxl}.

One  class of  gravitational modification with interesting applications is
mimetic gravity \cite{Chamseddine:2013kea,Chamseddine:2014vna}, which can be
obtained   from general relativity through the  isolation of the conformal
degree of freedom in a covariant way, by applying the a re-parametrization of
the
physical metric  in
terms of    a mimetic field  and an auxiliary metric. In this way, the
field equations exhibit an additional term arising from the
mimetic field. Since in a cosmological framework this term may be considered to
correspond to a dust fluid component, the theory could be applied to describe
 cold dark matter in a ``mimetic'' way. Nevertheless, mimetic gravity can be
extended in many ways, interpreted as a modification of gravity
\cite{Chamseddine:2014vna,Nojiri:2014zqa,Leon:2014yua, Nashed:2016tbj,
Momeni:2015gka,Matsumoto:2015wja,Chamseddine:2016uef,Dutta:2017fjw,Nashed:2011fg,
Chamseddine:2016ktu,Vagnozzi:2017ilo, Casalino:2018tcd,Nashed:2021sji,
Casalino:2018wnc, Abbassi:2018ywq,Zhong:2018tqn,Nashed:2020kjh,
 Odintsov:2015wwp,
Nojiri:2016vhu,Sadeghnezhad:2017hmr,Gorji:2019ttx,Gorji:2018okn, ElHanafy:2017sih,
Bouhmadi-Lopez:2017lbx,Gorji:2017cai,Firouzjahi:2018xob,Chamseddine:2019bcn,
Nashed:2021pkc} (for a
review see  \citep{Sebastiani:2016ras}).
Since mimetic modified gravity have many interesting applications at the
cosmological framework (among them the ability to alleviate the cosmological
tensions \cite{Abdalla:2022yfr}, and to track possible quantum-related defects
\cite{Addazi:2021xuf}), an amount of research has been devoted to the
investigation of the spherically symmetric solutions too
\cite{Deruelle:2014zza,Myrzakulov:2015sea,Myrzakulov:2015kda,Odintsov:2015cwa,
Nojiri:2017ygt,Odintsov:2018ggm,Oikonomou:2015lgy,Gorji:2020ten,
Nashed:2018qag,Chen:2017ify,Nashed:2018aai,
Nashed:2018urj,BenAchour:2017ivq,Zheng:2017qfs,Shen:2019nyp,
Sheykhi:2020fqf,Nashed:2021ctg}.

 Mimetic theory does not exhibit any difference from general relativity in
flat spacetime. Therefore, we will test the mimetic theory in the frame of  stellar structure models   using two equations of states and  radial metric potential $g_{rr}$  and confront  the output results with GR. Additionally,  we are interested in extracting new
anisotropic star solutions in mimetic gravity by imposing the Tolman-Finch-Skea
metric \cite{Finch_1989}, and in particular to examine the stability of the
solutions as well as the behavior of the anisotropy. The plan of the work is as
follows. In Section \ref{S2} we briefly review mimetic gravity, presenting the
field equations. In Section \ref{S3} we extract new classes of anisotropic
solutions. Then in Section \ref{data} we use the data from the 4U
1608-52 pulsar in order to investigate numerically the features of the obtained
anisotropic stars, namely the profile of the energy density, as well as the
radial and tangential pressures, the anisotropy factor, and the radial and
tangential equation-of-state
 parameters. In Section \ref{stability} we study the stability of the
solutions, applying the  Tolman-Oppenheimer-Volkoff (TOV) equation, the
adiabatic index, and we examine the static case. Finally, in Section \ref{S5}
we summarize the obtained results.

\section{Mimetic gravity}\label{S2}

  In this section we briefly present  mimetic gravity. Starting from general
relativity and    parametrizing the physical metric $g_{\mu
\nu}$ introducing an auxiliary metric $\bar{g}_{\mu \nu}$ and a mimetic
field $\phi$,
we acquire \cite{Chamseddine:2013kea}
\begin{eqnarray}
g_{\mu \nu} = \bar{g}_{\mu \nu}\bar{g}^{\alpha
\beta}\partial_{\alpha}\phi\partial_{\beta}\phi \, ,
\label{mimeticform}
\end{eqnarray}
 which implies that  the physical metric remains
invariant under
conformal transformations of the auxiliary metric. Hence, one can easily extract
the expression
\begin{eqnarray}
g^{\mu \nu}\partial_{\mu}\phi\partial_{\nu}\phi ={-} 1 \, ,
\label{cond11}
\end{eqnarray}
which can then be applied as   a Lagrange multiplier in an extended action,
namely
\citep{Chamseddine:2014vna}
\begin{eqnarray}
S = \int d^4x \sqrt{-g} \left [ \frac{R}{2 \kappa^2} + \lambda \left(
\partial^\mu\phi\partial_\mu\phi {+}1 \right) + \mathcal{L}_m \right ] \, ,
\label{action111}
\end{eqnarray}
with $\kappa^2$ is the gravitational constant, $R$   the Ricci scalar, and
$\mathcal{L}_
m$   the usual   matter  Lagrangian, in units where $c=1$.

One can extract the field equations  by varying the action in terms of  the
physical metric, incorporating additionally  its dependence on the   mimetic
field as well as the auxiliary metric, resulting to
\begin{equation}
\frac{1}{\kappa^2} G_{\mu\nu}=\lambda \partial_\mu \phi\partial_\nu
\phi+ T_{\mu \nu}, \label{eq000}
\end{equation}
with $G_{\mu\nu}$   the Einstein tensor and $T_{\mu \nu}$   the
standard  matter energy-momentum
tensor. Moreover,  variation with respect to the
Lagrange
multiplier  leads to the  condition  \eqref{cond11}.
 Taking the trace of equation (\ref{eq000}) we find the Lagrange
multiplier to be
$
\lambda=  G/\kappa^2-T $,
where $G$ and $T$ are the traces of the Einstein tensor and the matter
energy-momentum
tensor respectively. Varying      \eqref{action111} with
respect to
the mimetic scalar field leads to
\begin{equation}
\nabla^{\mu}\left(\lambda\partial_{\mu}\phi \right) =0
\label{eq:KG_eq11}.
\end{equation}
Lastly, the above equations can be elaborated in a more compact form, namely
\begin{eqnarray} \label{fe3}
&& G_\mu{}^\nu-(G-T)\partial_\mu \phi  \partial^\nu \phi=\kappa^2
T_\mu{}^\nu\,,  \\&&
 \nabla_\mu\left[(G-T)\partial^\mu \phi\right]=0\,.\label{fe3b}
\end{eqnarray}

In this work we
consider  $T_\mu{}^\nu$ to correspond to an anisotropic fluid, namely we impose
the form
\begin{eqnarray}
&&T_\mu{}^\nu{}=(p_t+\rho)u_\mu u^\nu-p_t\delta_\mu{}^\nu+(p_r-p_t)\xi_\mu
\xi^\nu\, ,
\end{eqnarray}
where $u_\mu$ is the time-like vector defined as $u^\mu=[1,0,0,0]$ and $\xi_\mu$
is the space-like  unit radial  vector defined as
$\xi^\mu=[0,1,0,0]$, such that $u^\mu u_\mu=-1$ and $\xi^\mu\xi_\mu=1$\footnote{In the present study we assume the Lagrangian multiplier to has a unite value.}.

\section{Novel classes of anisotropic solutions }\label{S3}

We are interested in extracting new classes of spherically symmetric
solutions of  the field equations (\ref{fe3}),(\ref{fe3b}). For this purpose,
we introduce
the metric
{
\begin{equation}\label{met}
ds^{2}=\left[h(r)h_1(r)\right]dt^{2}-h_1(r)dr^{2}-r^2\left(d\theta^{2}
+r^2\sin^2\theta d\phi^{2}\right)\,,
\end{equation}}
where $h(r)$ and $h_1(r)$ are the two metric functions. Under this metric, the
field equations (\ref{fe3}),(\ref{fe3b})  give rise to  the
following non-linear differential equations:
\begin{eqnarray}
&&
\!\!\!\!\!\!\!\!\!\!\!\!\!\!\!\!\!\!\!\!
{\frac {h'_1 r+
 h_1{}^{2}-h_1}{r^2\,h_1{}^{2}}}
=\kappa^2\rho\,,\label{fe} \\
&&\!\!\!\!\!\!\!\!\!\!\!\!\!\!\!\!\!\!\!\!
\frac {1}{2 h_1{}^{4}{r}^{2}h
^{2}}\Bigg\{2\,h_1{}^ {4} h^{2}-2\,  h'_1 r
h^{2}h_1{}^{2}-2\, h'  rh h_1{}^{3}-2\, h^{2}h_1{}^{3}- 2\, \phi'^{2}{r}^{2}
h^{2}h_1{}^{3}(\rho -p_r -2p_t)\nonumber\\
&& \ \ \ \ \,
+4\, \phi'^{2}( h^{2} h_1{}^{3}-h_1{}^{2}
h^{2}-r h' h h_1{}^{2})
-2\, \phi'^{2}{r}^{2} h'' h
h_1{}^{2}+2\,\phi'^{2}{r}^{2}h'_1{}^{2}
h^{2}
\nonumber\\
&& \ \ \ \ \, -\phi'^{2}{r}^ {2}h'   h'_1{}  hh'_1{}
-2\,\phi'^{2}{r}^{2} h''_1{}
h^{2}h_1{}+
\phi'^{2}{r}^{2}h'^{2}h_1{}^{2}\Bigg\}=-\kappa^2\,p_r\,,\label{fe1}\\
&&
\!\!\!\!\!\!\!\!\!\!\!\!\!\!\!\!\!\!\!\!
\frac {2\,r
 h'_1{}^{2}h^{2}-2\,h' h
   h_1{}^{2}-2\,r
 h''  h h_1{}^{2}-r h' h'_1{} h h_1{}-2\,r\,h''_1{}h^{2} h_1{} +r h'^{2}
h_1{}^{2}}{4r h^{2}
  h_1{}^{3}}=-\kappa^2\,p_t\label{fe2}\,,
\end{eqnarray}
where   $\phi\equiv \phi(r)$  and with primes denoting derivatives with
respect to $r$.
The system  (\ref{fe})-(\ref{fe2}) consists of  three
independent equations for six unknown  functions;  $h$, $h_1$, $\rho$, $p_r$,
$p_t$, and $\phi$.  Hence, we require to impose three extra conditions. { We will try to solve the above system assuming the two equations of state: \[p_r =\omega_1\rho\,,\]  and \[p_T =\omega_2\rho\,,\] and by imposing the condition $\phi=\int \sqrt{-g_{rr}}dr=\int \sqrt{h_1}dr$ we get the following system:
\begin{eqnarray}\label{ref1}
&&-{\frac {-h h'_1 r- h' rh_1  +h  h_1{}^{2}-h h_1  }{ h_1{}^{2}h  {r}^{2}}}=-{\frac{ \omega_1\,\left(  h'_1r+ h_1{}^{2}-h_1   \right) }{  h_1  ^{2}{r}^{2}}}\,,\nonumber\\
&&\frac {-2\,r h''_1  h^{2}h_1 -2\,r h'' h  h_1 {}^{2}+r h'^{2} h_1{}^{2}-h h_1  \left( 2\,h_1 + h'_1   r \right) h' +2\,r h'_1{}^{2} h^{2}}{4r  h^{2} h_1{^{3}}}=-{\frac {\omega_2\, \left( h'_1   r+ h_1{}^{2}-h_1 \right) }{  h_1{}^{2}{r}^{2}}}\,,\nonumber\\
&&
\end{eqnarray}
where $\omega_1$ and $\omega_2$ are the parameters characterizing the fluid.
The above differential equations, \eqref{ref1} have no analytical solution except for the case of dust, i.e., $\omega_1=\omega_2=0$, the vacuum case otherwise, we cannot find any analytical solution that can extract from it any physics.

Other way to solve equations (\ref{fe})-(\ref{fe2}) is to assume the
expression of the metric potential $g_{rr}$, i.e.  $h_1$, as:}
\begin{equation}\label{sol1}
 h_1=\left(1+\frac{r^2}{k}\right)^s\,,
\end{equation}
where $s$ is a real parameter  and $k$  is a constant.  If  $s=0$ then Eq.
(\ref{fe}) yields $\rho=0$ which is not physically interesting.
The gravitational potential  (\ref{sol1}) is  well-behaved and finite when
$r\rightarrow 0$.  For $s=1$  the metric potential reduces  to the well known
Tolman-Finch-Skea potential \cite{Finch_1989}, which  has been applied
 to model compact stars by using a proper choice of the
radial pressure $p_r$,  compatible with
observational data \cite{Pandya et al.(2015)}.

At this stage it is important to introduce the anisotropy parameter $\Delta$,
defined as \cite{Nashed:2021pkc}
\begin{eqnarray}
  \Delta=\kappa^2(p_t-p_r),
\end{eqnarray}
which quantifies the amount of anisotropy present in the star, being zero in
the isotropic case. Inserting (\ref{sol1}) into  (\ref{fe})-(\ref{fe2}),   we
can obtain the expression of anisotropy  as
{\small{
\begin{eqnarray}\label{d1}
 &&
 \!\!\!\!\!\!\!\!\!\!\!\!\!\!\!\!
 \Delta=\frac {1}{r^2}-\frac{ \left( 1+3s \right)
{r}^{4}{k}^{2\,s}+{r}^{2}
\left( 2+s \right) {k}^{2+2s}+{k}^{4+
2\,s}  }{{r}^{2}\left( {k}^{2
}+{r}^{2} \right) ^{2+s}}
\nonumber\\
 &&\!\!\!\!\!\!\!+\frac {2 {r}^{2}
 \left\{ 2rh h'' \!+\! h'  \left[h  ( s\!-\!2 ) -
rh'   \right]  \right\} {k}^{2\!+2\!s}+ (2 rh  h''\!-\!2h h' \! - \!h'^{2}r )
{k}^{4\!+\!2s}+{r}^{4}{k}^{2\,s} \left\{
2rh h''  +
 \left[2h
 ( s\!-\!1 )  \!-\!rh'\right]h'
 \right\} }{4r  \left( {k}^{2}\!+\!{r}^{2} \right) ^{s+2} h^{2}}
.
\end{eqnarray}}}
Imposing the condition
{\small{
\begin{equation}\label{sold} \frac {2 {r}^{2}
 \left\{ 2\,rh h'' + h'  \left[h  ( s\!-\!2 ) -
rh'   \right]  \right\} {k}^{2+2\,s}+ (2 rh  h''-2h h'  - h'^{2}r )
{k}^{4+2\,s}+{r}^{4}{k}^{2\,s} \left\{
2rh h''  +
 \left[2h
 ( s\!-\!1 )  -rh'\right]h'
 \right\} }{4r  \left( {k}^{2}+{r}^{2} \right) ^{s+2} h^{2}}=0
,
\end{equation}}}
we find
\begin{equation}\label{sol2}
 h  =\frac { \left\{c_2 \left( s-
2 \right) -c_1\,{2}^{s/2}
 \left( {k}^{2}+{r}^{2} \right)  \left[\left( {k}^{2}+{r}^{2}
 \right)  \left( 2+s \right)  \right] ^{-s/2}  \right\} ^{2}}{4 \left( s-2
\right) ^{2}}\,,
\end{equation}
where $c_1$ and $c_2$ are integration constants  which can be determined from
the matching conditions.
Substituting (\ref{sol1}) and (\ref{sol2}) into  (\ref{fe})-(\ref{fe2}) we
obtain the energy density, as well as the  radial and tangential pressures,
respectively as:
{\small{
\begin{eqnarray}
\label{psol}
&&
\!\!\!\!
\kappa^2 \rho(r)=\frac{1 - \left(1 + \frac{r^2}{k^2}\right)^{-s}}{r^2} +
\frac{2 s \left(1 + \frac{r^2}{k^2}\right)^{-1 -
s}}{k^2}\label{den}\,,\nonumber\\
&&\!\!\!\!
\kappa^2 p_r (r)
=\frac{ c_2\left\{ ( k^{2}\!
+\!{r}^{2} ) ^{s+1}\!-\! k^{2s}\left[( 1\!+\!2s ) {r}^{2}\!+\!k^2\right]
\right\}  (s\! -\!2 )  \! -\!c_1 ( k^{2}\!+\!{r}^{2} )^{1-s/2}
\left[
  (k^{2}\!+\!{r}^{2} ) ^{s+1}\!-\!k^{2s}({k}^{2}\!-\!5{r}^{2})
\right] {2}^{s/2} ( 2\!+\!s)^{-s/2}}{ r^2 \left[ c_1{
2}^{s/2} ( k^2\!+\!{r}^{2} )^{1-s/2}
( 2\!+\!s )^{-s/2}-c_2
 ( s\!-\!2 )  \right] (k^{2}\!+\!{r}^{2})^{1+s}}\,,\nonumber\\
&& \!\!\!\!\kappa ^2 p_t (r) =-\frac{k^{2s}\left\{ c_2s\left( k^{2}
-{r}^{2}\right)  \left(s -2 \right) +c_1 \left[k^2(s-4)+r^2(3s-4)\right]\left(
k^{2}+{r}^{2} \right)^{1-s/2}  {2}^{s/2} \left(
2+s\right)^{-s/2}\right\}}{\left[ c_1{
2}^{s/2} \left( k^{2}+{r}^{2} \right)^{1-s/2}  \left( 2+s \right)^{-s/2}-c_2
 \left( s-2 \right)  \right] \left(k^{2}+{r}^{2} \right)^{1+s}}\,.
\end{eqnarray}}}
Finally, the metric solution in the interior of the star is written as
\begin{equation}
ds_{-}^2=\frac{\left[ c_1{2}^{s/2} \left( {k}^{2}\!+\!{r}^{2} \right)^{1-s/2}
 \left(s+2 \right) ^
{-s/2}-c_2 \left( s-2 \right)  \right]^{2} \left(k^2+
{r}^{2}\right) ^{s}}{4k^{2s}\left(s -2 \right) ^2}dt^2-\left(1 +
\frac{r^2}{k^2}\right)^{s} dr^2-r^2(d\theta^2+\sin^2 \theta
d\phi^2).\label{interiorsolution}
 \end{equation}
 We stress here that according to (\ref{cond11}) the solution for the mimetic
field can be obtained as {$\phi(r)=\int dr\sqrt{g_{rr}} =
 {}_2F_1\left(\frac{1}{2}, -s; \frac{3}{2}; -\frac{r^2}{k^2}\right)r $}, with
$_2F_1$ the hypergeometric function. Hence, this solution does not include
general relativity result as a limit, since $\phi(r)$ is not constant, namely
it is a novel class of solutions.

The constraint  (\ref{sold}) is important since in this case the anisotropy
parameter does not directly depend on $h$, and
it acquires   the simple
 form
\begin{eqnarray}\label{d2}
 \Delta=\frac {1}{r^2}-\frac{ \left( 1+3\,s \right) {r}^{4}{k}^{2\,s}+{r}^{2}
\left( 2+s \right) {k}^{2+2\,s}+{k}^{4+2\,s} }{{r}^{
2}\left( {k}^{2}+{r}^{2} \right) ^{2+s}}\,.
\end{eqnarray}
This form  has the expected properties to    vanish
at the center of the star, i.e $\Delta(r\rightarrow0)=0$,  it has no
singularities, and  has a positive value  inside the
star
\cite{Dey:2020fxm,Maharaj:2014vva,Murad:2014oua}.
Note that  the anisotropic force, defined as
$\frac{2\Delta}{r}$,  is attractive for $p_r-p_t>0$ and  repulsive
 for $p_r-p_t<0$.

 From the above expressions we deduce that
 $\rho$, $p_r$,
$p_t$ are  well-defined at the center of the star, regular and
singularity-free.
In particular we find
\begin{eqnarray}
&&\rho(r\rightarrow 0)=\frac{3 s}{\kappa^2 k^2},\label{rho2}\\
&&p_r(r\rightarrow 0)=p_t(r\rightarrow 0)=\frac{sc_2\, \left( s-2\right)
{k}^{s-2}
 \left( 2+s \right)   ^{s/2}+ c_1\,{2}^{s/2}
 \left(s -4 \right) }{ \kappa^2\left[   c_2\, \left( s-2
 \right) {k}^{s}
 \left( 2+s \right)   ^{s/2}-c_1\,{2}^{s/2}{k}^{2}  \right] },\label{pr2}
\end{eqnarray}
and thus   (\ref{rho2}) implies that in physical cases $s>0$.
 Additionally, note that for a large $s$-region  $\rho$, $p_r$,
$p_t$   are non-negative,
regular and  singularity-free.
 Finally, note that calculating the gradients of the  density and pressures
from
the above expressions,
namely  $\frac{d\rho}{dr}$,
$\frac{dp_r}{dr}$,  and $\frac{dp_t}{dr}$, we can immediately verify that they
are negative, as required, and in particular
 $\rho$ is   finite and
monotonically decreasing towards  the boundary.

 We can now introduce the radial and tangential equation-of-state
parameters $w_t$ and  $w_r$  as
  \begin{eqnarray} w_t\equiv\frac{p_t}{\rho}\,,\qquad \qquad
w_r\equiv\frac{p_r}{\rho}\,,
\end{eqnarray}
while in cases of anisotropic objects it is convenient to introduce also the
average equation-of-state
parameter
 \begin{eqnarray}
 w_{av}\equiv\frac{p_r+2p_t}{\rho}.
\end{eqnarray}
Furthermore, we can calculate the radial and tangential     sound speeds,
$v_r{}^2=\frac{dp_r}{d\rho}$ and $v_t{}^2=\frac{dp_t}{d\rho}$ respectively,
which are given in   Appendix \ref{AppA}.  Finally, the mass contained within
radius $r$
of the sphere
is defined as:
\begin{align}\label{mas}
M(r)={\int_0}^r \rho(\xi)\xi^2 d\xi\,.\end{align}
Inserting (\ref{psol}) into  (\ref{mas}) we acquire
\begin{align}\label{mas1}
M(r)=\frac{ r\left[  \left( {\frac {{k}^{2}+{r}^{2}}{{k}^{2}}} \right) ^{s}-1
\right]}{2 \left( {\frac {{k}^{2}+{r}^{2}}{{k}^{2}}} \right) ^
{s}}\,.\end{align}
Thus, we can now introduce the compactness parameter of a spherically symmetric
source with radius $r$ as \cite{NewtonSingh:2019bbm}
\begin{eqnarray}\label{gm1}
&&m(r)=\frac{2M(r)}{r}=\frac{\left( {\frac {{k}^{2}+{r}^{2}}{{k}^{2}}} \right)
^{s}-1}{ \left( {\frac {{k}^{2}+{r}^{2}}{{k}^{2}}} \right) ^{s} }\,.
\end{eqnarray}

We proceed by determining  the constants $c_1$, $c_2$ and $k$. To achieve
this we match the interior solution (\ref{psol})
and the interior metric (\ref{interiorsolution}),
 with  the exterior
Schwarzschild solution\footnote{ We have shown in \cite{Nashed:2021ctg} that the only vacuum spherically symmetric solution, in the frame of mimetic gravitational theory, is the Schwarzschild one.}
\begin{eqnarray}
ds_+^{2}=\left(1-\frac{2M}{r}\right)dt^{2}-\left(1-\frac{2M}{r}\right)^{-1}dr^{2
}-r^{2}\left(d\theta^{2}+\sin^{2}\theta d\phi^{2}\right)\,,
\end{eqnarray}
for $r>2M$,  and with $M$ the total mass of the compact star
(note that the metric (\ref{met})
reproduces the Schwarzschild solution in vacuum).
The junction condition of the metric potentials across the boundary is given by
the first fundamental form, namely at the surface $r\rightarrow l$,
as $g_{rr}^+=g_{rr}^-$, and $g_{tt}^+=g_{tt}^-$,
and the second fundamental form implies
$
p_r(r\rightarrow l)=0$. These conditions lead to
   \begin{eqnarray}
&&
\!\!\!\!\!\!\!\!\!\!\!\!\!\!\!\!\!\!\!
c_1=\pm l^{s-3}\left\{  \left( s+2 \right)   \left[  \left( {\frac
{l}{l-2
\,M}} \right) ^{{1/s}}-1 \right] ^{-1} \right\} ^{s/2}
\nonumber\\
&&
\cdot\left\{
{2}^{-s/2}s  \left( 2M-l \right)  \left({\frac {l}{l-2M}} \right) ^{{\frac
{s-2}{2s}}}+ {2}
^{-s/2}\left[2Ms+\left(M -sl\right)  \right] \sqrt {{\frac {l}{l-2M}}}
 \right\}\,,
\label{b2}
 \\
&&
\!\!\!\!\!\!\!\!\!\!\!\!\!\!\!\!\!\!\!
c_2=\pm\frac{\left[\frac{ 2 \left( 2l-5M
\right)}{  \left( {\frac {l}{l-2\,M}}
\right) ^{{s}}}-4(l-2M) \right]}{ {l} \left[  \left({
\frac {l}{l-2M}} \right) ^{{s}}-1 \right] \left(s -2
 \right)}
\,,\label{b3}\\
&&
\!\!\!\!\!\!\!\!\!\!\!\!\!\!\!\!\!\!\!
k=\pm l^2\sqrt { \left[  \left({\frac{l-2\,M}
{l}} \right) ^{{s}}-1
 \right] ^{-1}}\,. \label{kb3}
 \end{eqnarray}

 We close this section by  analyzing the properties of the metric
solution (\ref{interiorsolution}).
   Equations~(\ref{sol1}) and (\ref{sol2}) imply that
the metric potentials $g_{tt}$ and $g_{rr}$ at the center of the star $r=0$
become
\begin{eqnarray}
&&\!\!\!\!\!\!\!\!\!\!\!\!\!\!\!\!\!
g_{rr}|_{r\rightarrow0}=1,\\
&&\!\!\!\!\!\!\!\!\!\!\!\!\!\!\!\!\!
     g_{tt}|_{r\rightarrow0}=\frac
{c_2^{2} \left( s\!-\!2 \right) ^{2}\left[
{k}^{2} ( 2\!+\!s )  \right]^{3s/2}+c_1
 \left\{ {2}^{s/2}\sqrt { \left[ {k}^{2} ( 2\!+\!s )  \right] ^{s}}{k}^{2
}c_1-2c_2  \left[ {k}^{2} ( 2\!+\!s
)  \right] ^{s} (s\!-\!2 )  \right\} {2}^{s/2}{k}^{2
}}{ 4(s\! -\!2 ) ^{2} \left\{  \left[ {k}^{2} ( 2\!+\!s
)  \right] ^{s} \right\} ^{3/2}}
 ,
 \end{eqnarray}
  and thus  the star  is free from a singularity at the center.

 \section{Physical features of the   solutions }
\label{data}

 In this section we proceed to the investigation of the physical features of
the  obtained anisotropic
solutions.
Any  physical viable stellar model must  satisfy the following
conditions throughout the stellar configurations:
\begin{itemize}
 \item

  The   metric potentials, and all components of the energy-momentum tensor,
  must  be well-defined and regular  throughout
the interior of the star.
 \item  The density  must  be finite and positive in
the interior of the star,   and decrease monotonically
 toward the boundary.
 \item  The radial and the tangential pressures
must be positive inside the configuration of the fluid, and the derivatives
of the density and pressures must be
negative. Additionally, the radial pressure $p_r$ must vanish at the boundary
of the stellar model $r\rightarrow l$, however the tangential pressure $p_t$
does not need to be zero at the boundary. Finally, at the center of the star
the pressures should be equal, implying
 that the anisotropy  vanishes, namely $\Delta(r = 0) = 0$.

 \item  Any anisotropic fluid sphere must   fulfill  the   energy
conditions, namely the null energy condition (NEC): $p_t+\rho > 0$, $\rho>
0$, the  strong energy condition (SEC): $p_r+\rho > 0$, $p_t+\rho > 0$,
$\rho-p_r-2p_t > 0$, the weak energy condition (WEC): $p_r+\rho > 0$,
$\rho> 0$, and the dominant energy condition (DEC): $\rho\geq \lvert
p_r\lvert$ and $\rho\geq \lvert p_t\lvert$.

 \item  The interior metric potentials must join smoothly
with the Schwarzschild  exterior metric at the boundary.
 \item  For a stable configuration, the adiabatic index must be
greater than $\frac{4}{3}$.
 \item  The stability of the
anisotropic stars should satisfy $0>v_r{}^2-v_t{}^2>-1$ where $v_r^2$ and
$v_t^2$
are
the radial and tangential sound speed squares respectively
\cite{Herrera:1992lwz,Boehmer:2006ye}.

 \item  The causality condition must be
satisfied,  namely the sound speeds   must be sub-luminal, i.e.
$0\leq\frac{dp_r}{d\rho}\leq 1$, $0\leq
\frac{dp_t}{d\rho}\leq 1$.

 \item The surface redshift $Z_R$
  is defined as the   value of
  \begin{eqnarray}
  Z(r)=\frac{1}{-g_{00}(r)}-1
  \label{redshift0}
 \end{eqnarray}
 calculated at the surface of the star, namely
$Z_R=\frac{1}{-g_{00}(R)}-1$
  \cite{Buchdahl:1959zz},
where $g_{00}$ is the temporal component of the metric, and it must  obey
$Z_R\leq 2$.

\end{itemize}

Let us examine whether our solutions satisfy
the above necessary physical conditions. In order to proceed we need to give
numerical values to the model parameters $c_1$, $c_2$ and $k$.
This will be obtained by using as input values the mass and radius of the
pulsar \textrm{4U 1608-52},   estimated respectively as  $M =
1.57_{-0.29}^{+0.3}M_\odot$ and $l=9.8\pm1.8$ km
\cite{Gangopadhyay:2013gha,Das:2021qaq,Roupas:2020mvs}.   Inserting these
values into (\ref{b2})-(\ref{kb3}),   we find
\begin{eqnarray}
&&c_1\approx {\frac {2}{729}}\, \left[ {\frac {81(s+2)}{{ 2.1}^{{s}^{-1}}-1
}} \right] ^{s/2}2^{-s/2} \left[
5.3 s ({ 2.1}^{{\frac
{s-2}{2s}}}-1.2)  +
 3.4 \right]\,,\nonumber\\
 && c_2\approx\frac {2( 8.5- 6.2\cdot{ 2.1}^{{s}^{-1}})}
 {9 ( 2.1^{{s}^{-1}}-1)(s-2) }\,, \nonumber\\&&
 k\approx
9\,\sqrt { \left( {e^{ 0.7{n}^{-1}}}- 1 \right)
^{-1}},
\end{eqnarray}
where $(c_1)^{\frac{1}{s-2}}$ has units of km, $c_2$ is dimensionless and $k$
has units of km$^2$.
We mention that apart from 4U 1608-52 a similar analysis can be developed
for other pulsars, such as  4U
1724-207  and
J0030+0451, and for completeness we provide the corresponding parameter values
in   Appendix \ref{AppB}.
Adopting the above  constants, the physical quantities extracted in the
previous
section can be plotted.

\begin{figure}
\centering
 \includegraphics[scale=0.26]{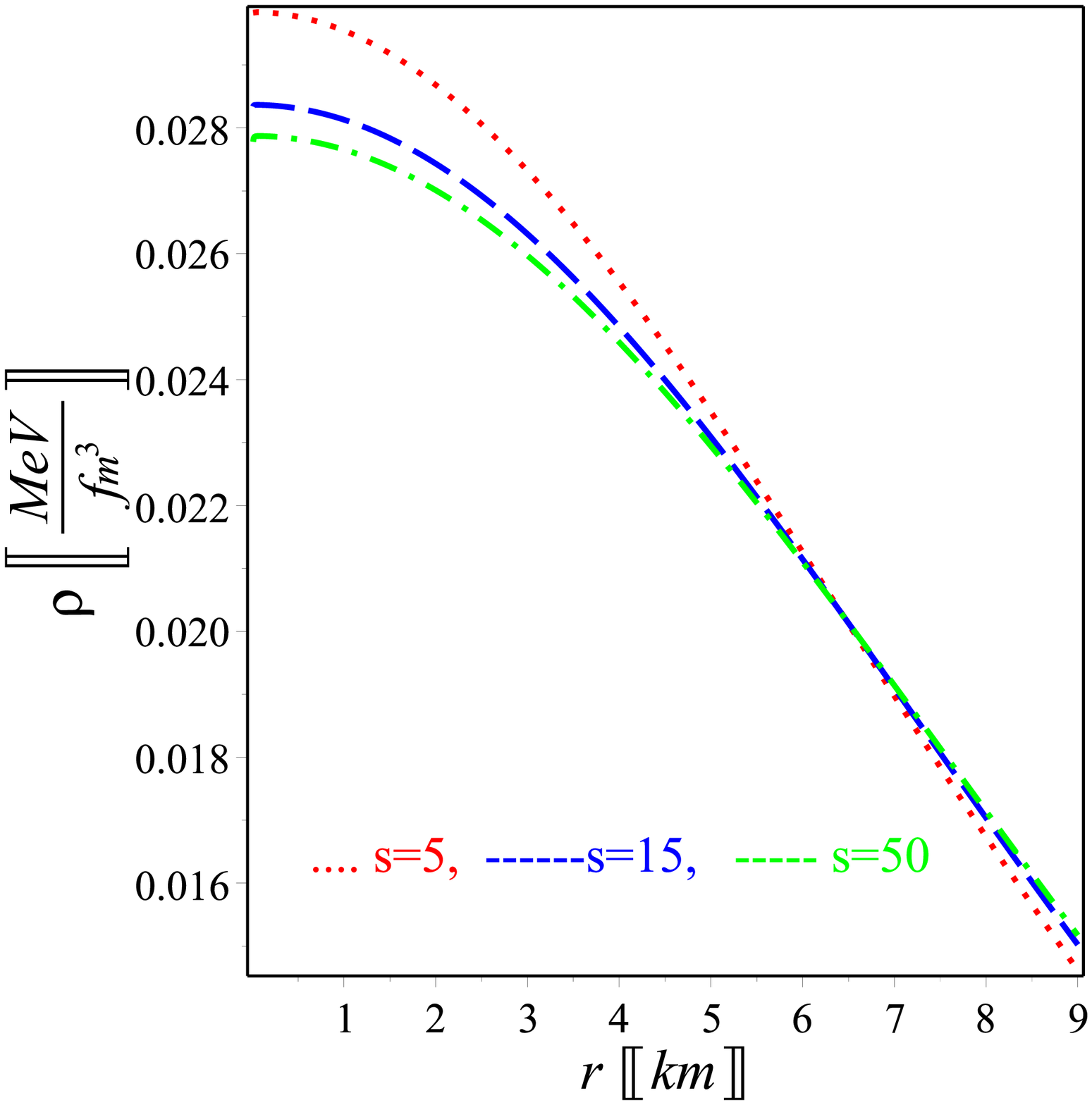}
\includegraphics[scale=.26]{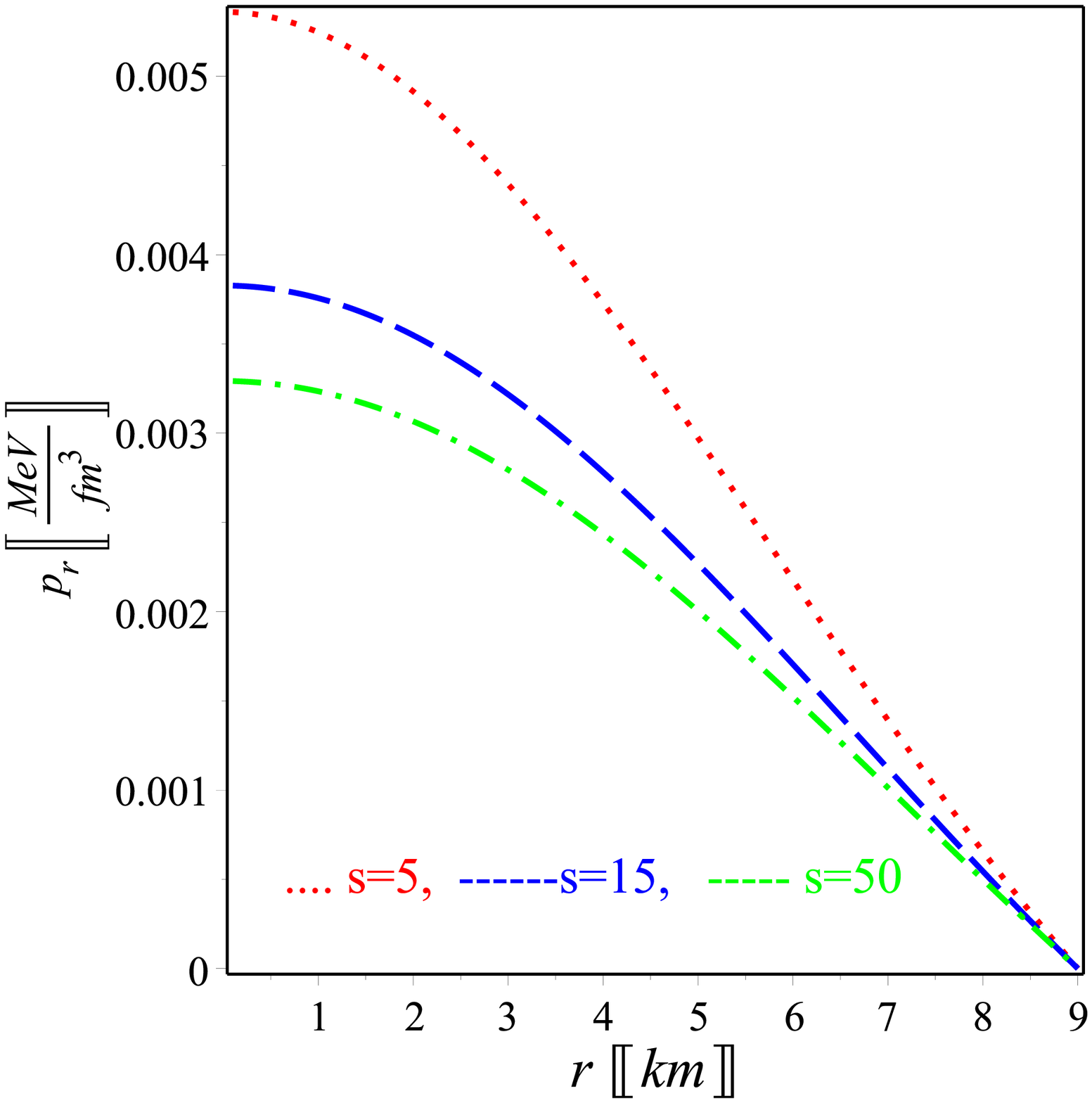}
\includegraphics[scale=.26]{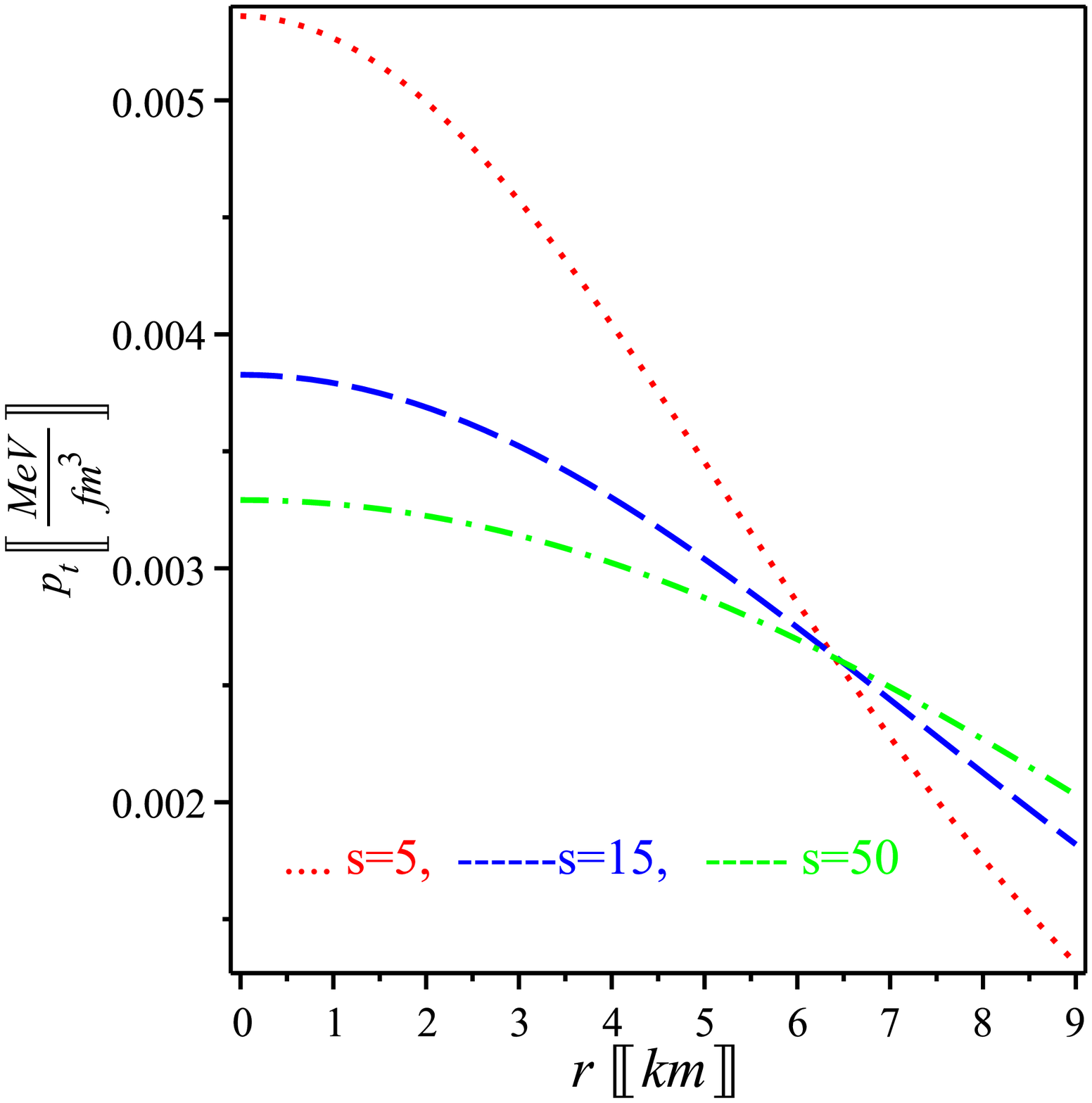}
\caption {\it{{ The energy density (left graph), the
radial pressure (middle graph) and the tangential pressure (right graph) of the
anisotropic star  solution
(\ref{psol}), as functions of the radial distance, for various values of the
parameter $s$, using the  4U 1608-52    mass and radius values. }}}
\label{Fig:1}
\end{figure}

The profiles of the energy density, radial and tangential pressures, given by
(\ref{psol}) are depicted in Fig.  \ref{Fig:1}. As we observe,
they are  well-defined at the center of the star, regular and
singularity-free, and they are positive and monotonically decreasing towards
the
boundary.

   \begin{figure}[ht]
\centering
\includegraphics[scale=.26]{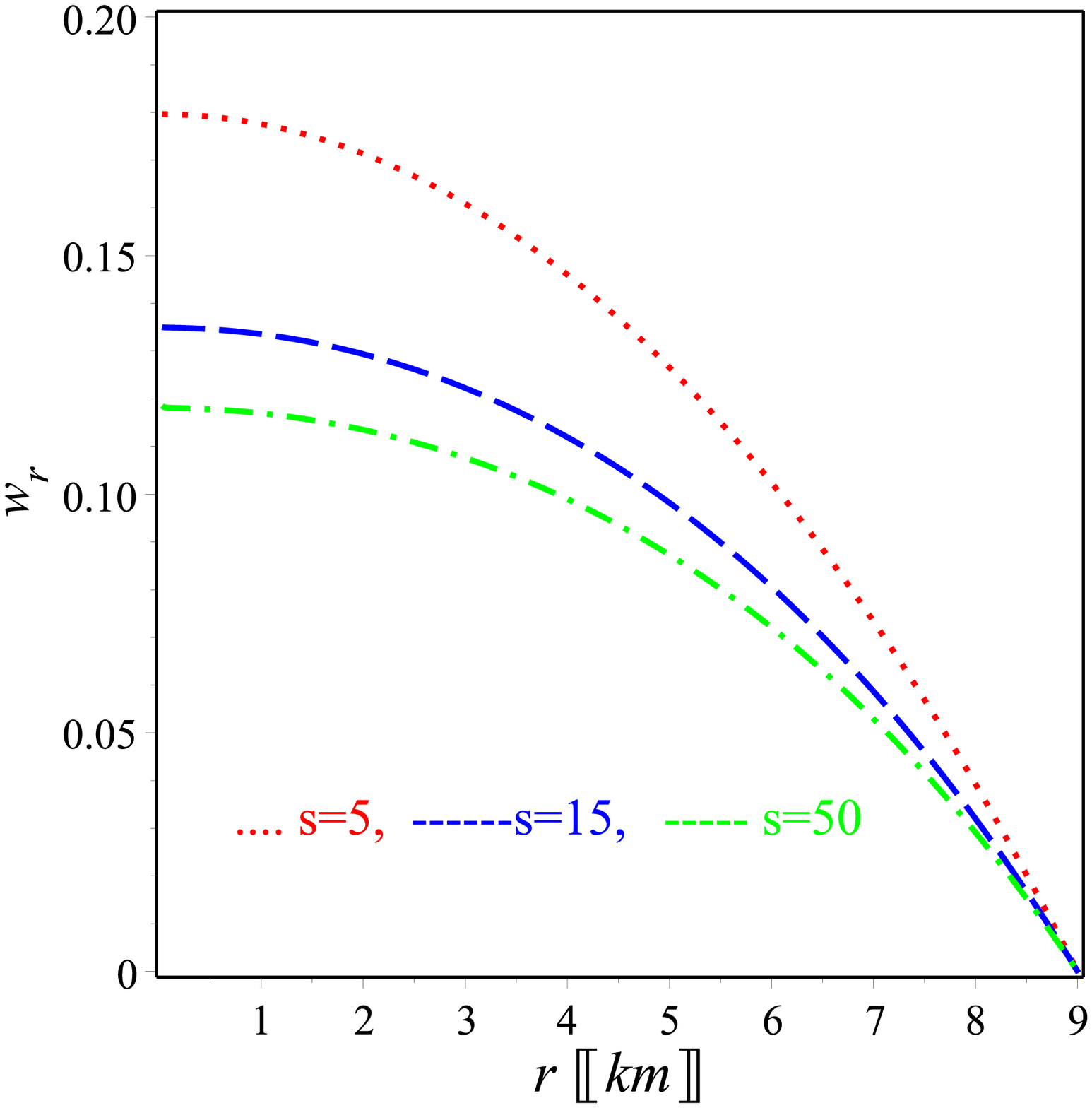}
\includegraphics[scale=0.26]{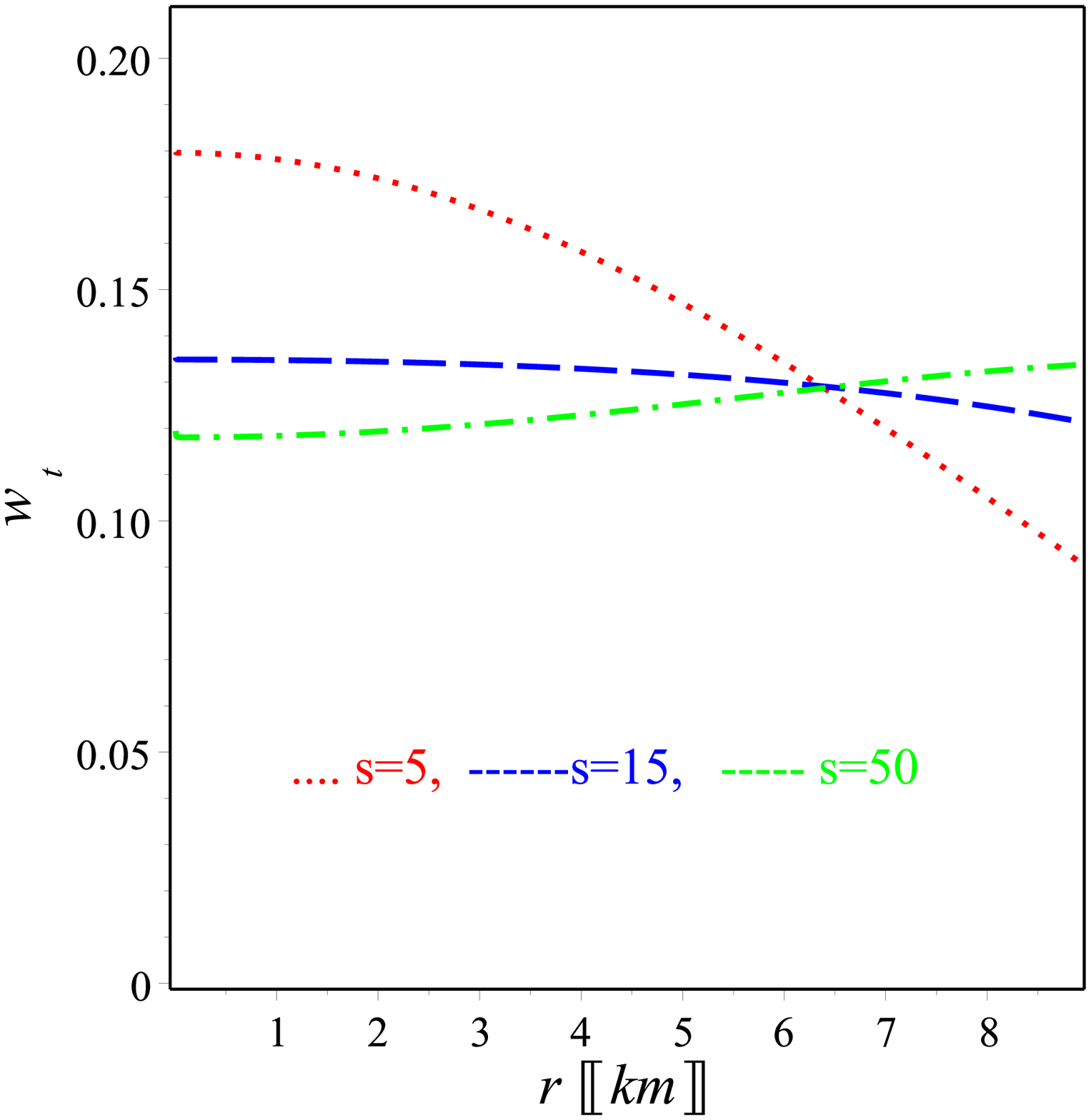}
\includegraphics[scale=.26]{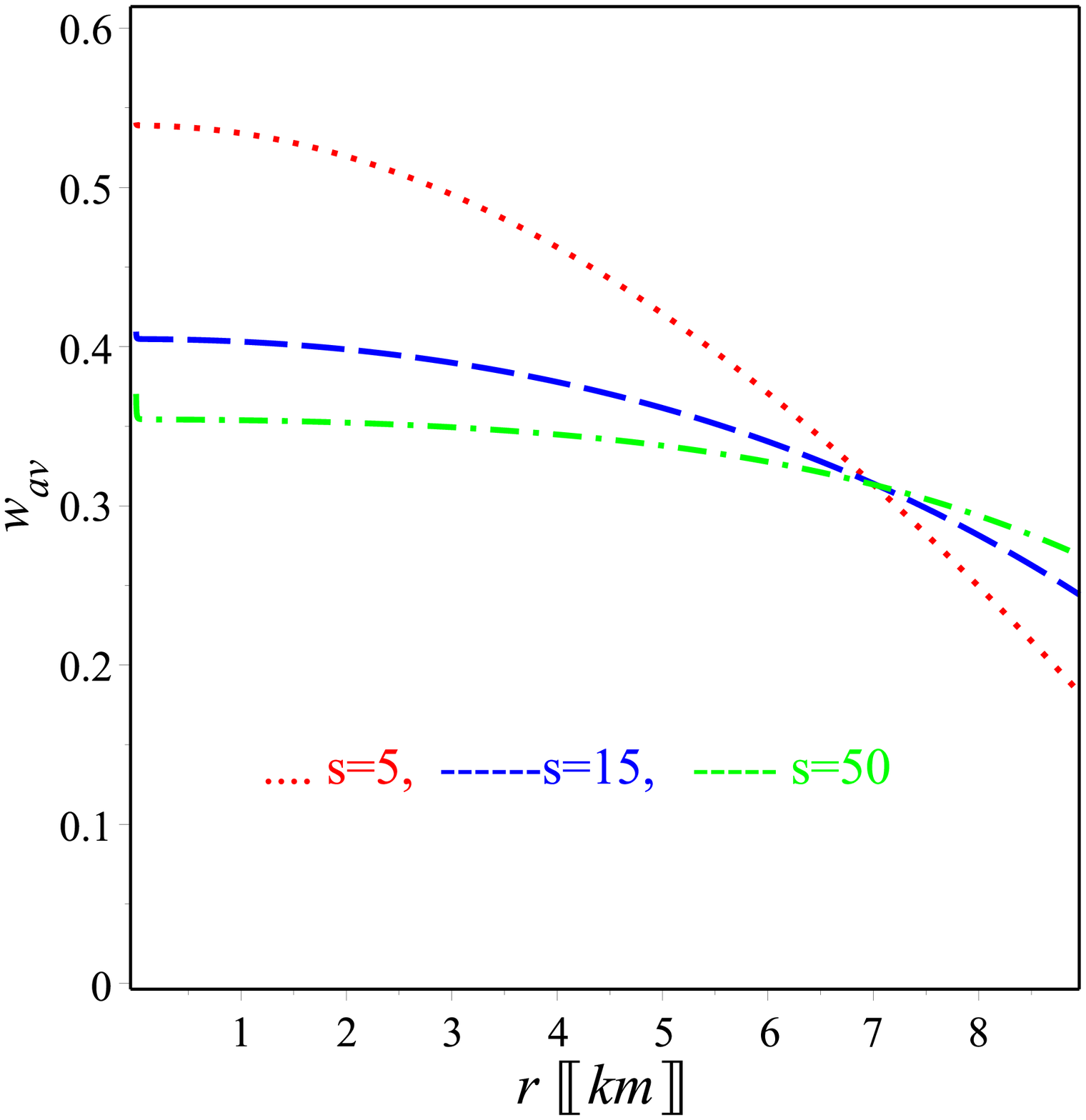}
\caption {\it{{The radial equation-of-state parameter $w_r$  (left graph), the
tangential equation-of-state parameter   $w_t$ (middle graph) and the
average equation-of-state parameter    (right graph) of the
anisotropic star  solution
(\ref{psol}), as functions of the radial distance, for various values of the
parameter $s$, using the  4U 1608-52
mass and radius values. }}}
\label{Fig:2}
\end{figure}

\begin{figure}
\centering
 \includegraphics[scale=.3]{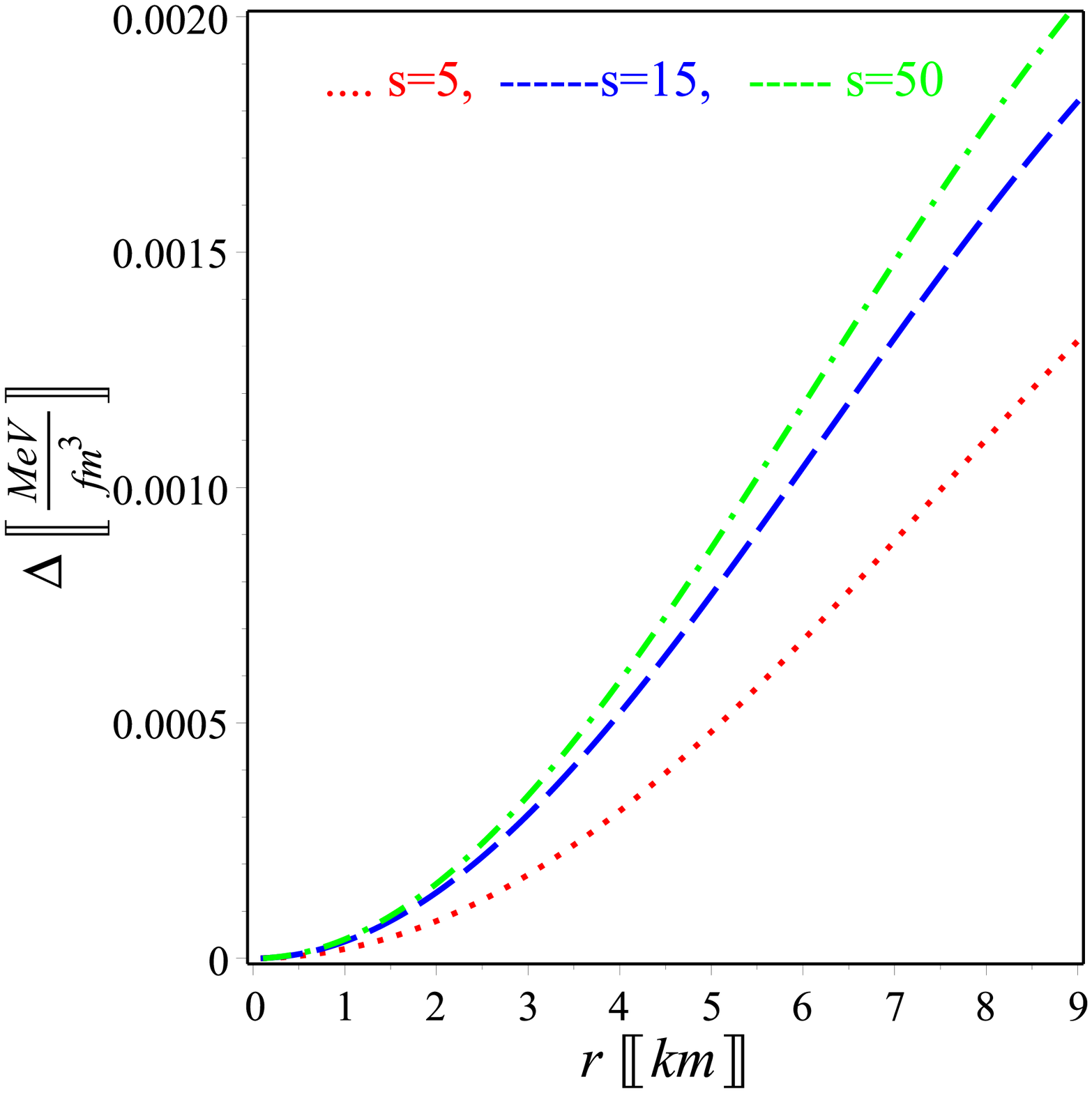}
 \includegraphics[scale=.3]{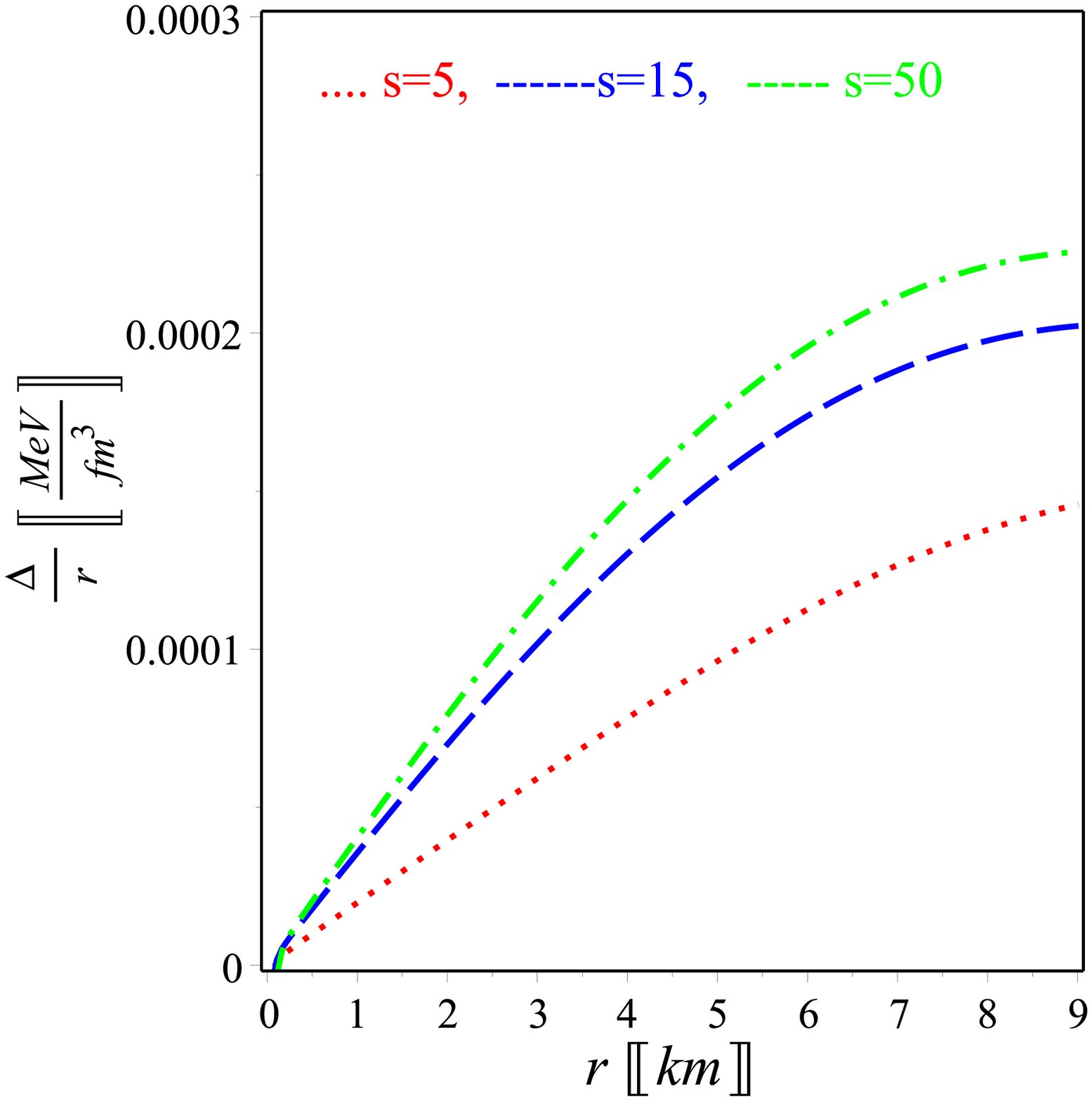}
\caption {\it{{The anisotropy  $\Delta$  (left graph), and the
  anisotropic force
$\frac{\Delta}{r}$  (right graph) of the
anisotropic star  solution
(\ref{psol}), as functions of the radial distance, for various values of the
parameter $s$, using the  4U 1608-52
mass and radius values. }}}
\label{Fig:2b}
\end{figure}

In Fig.   \ref{Fig:2}  we depict the radial, tangential, and average
equation-of-state
parameters,  $w_r$,  $w_t$ and  $w_{av}$ respectively. As we observe
   $w_r$ is monotonically increasing and $w_{av}$ is monotonically decreasing
with $r$, while
$w_t$ is monotonically decreasing for small $s$ and monotonically
increasing for
large $s$. Moreover, the values of $w_r$,  $w_t$ and  $w_{av}$   are
positive  and
lie in the interval  $\{w_t,\,\,w_r,\, w_{av}\}\in[0,\,\,1]$, which implies
that matter distribution is non-exotic in nature.
  Finally, in the left graph of Fig. \ref{Fig:2b}
  we depict the anisotropy $\Delta$, where we can see that
it is positive, it vanishes at the
center and it increases towards the surface of the star. Moreover, in the right
graph of Fig.
\ref{Fig:2b}    we depict the anisotropic force
$\frac{\Delta}{r}$, and the fact that it is positive implies that it is
repulsive.

\begin{figure}[ht]
\centering
\includegraphics[scale=0.26]{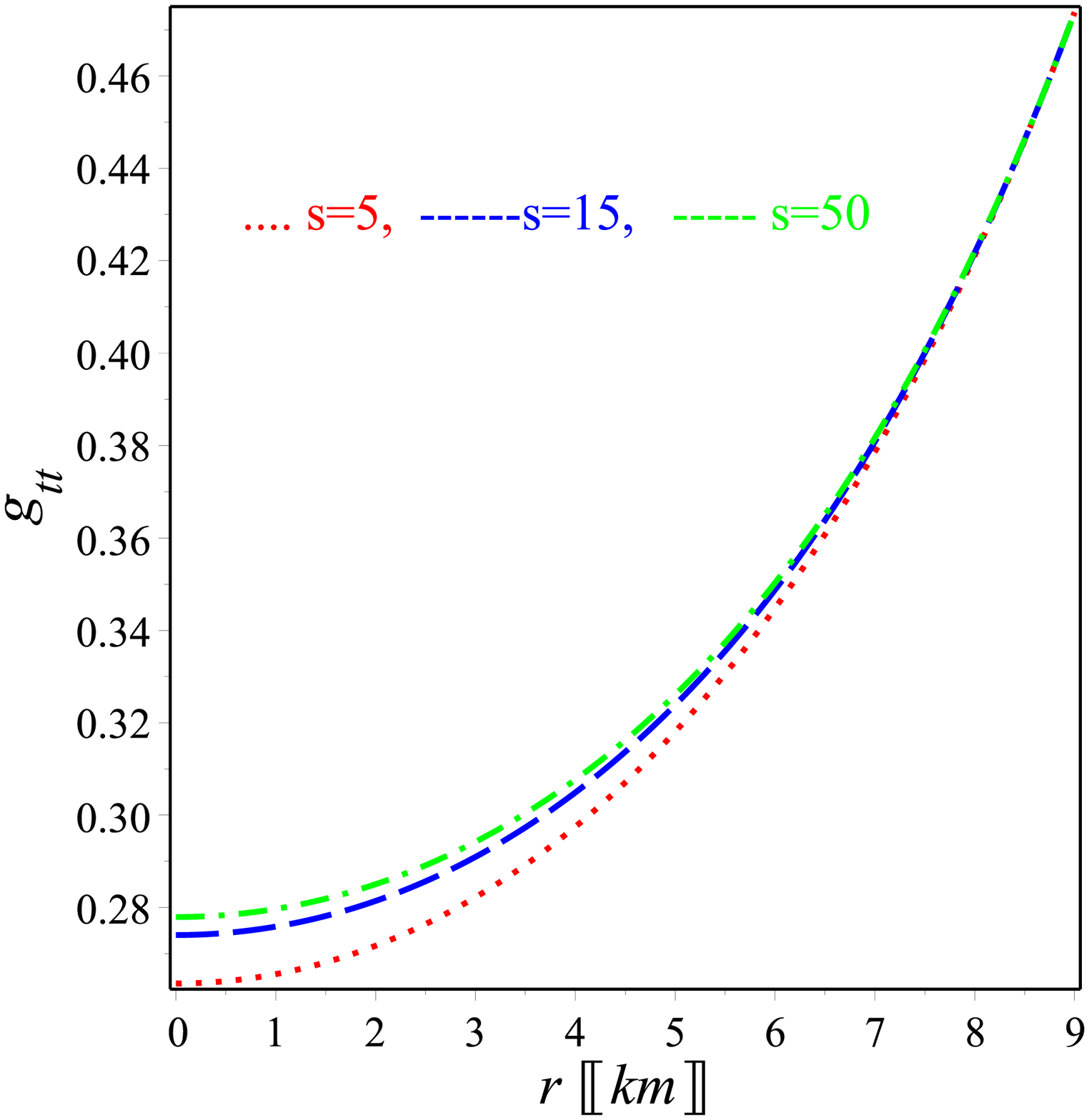}
\includegraphics[scale=0.26]{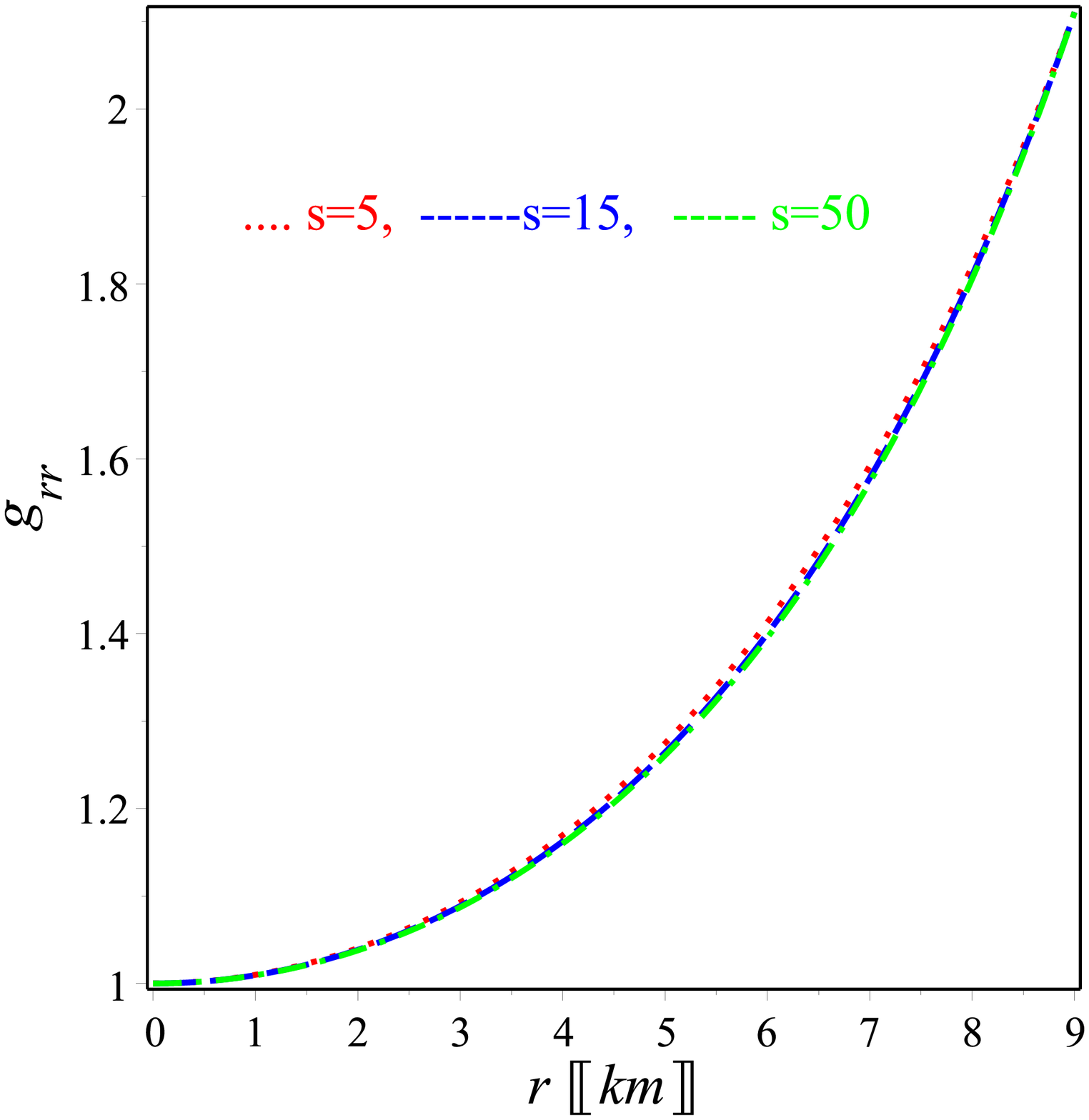}
\includegraphics[scale=.26]{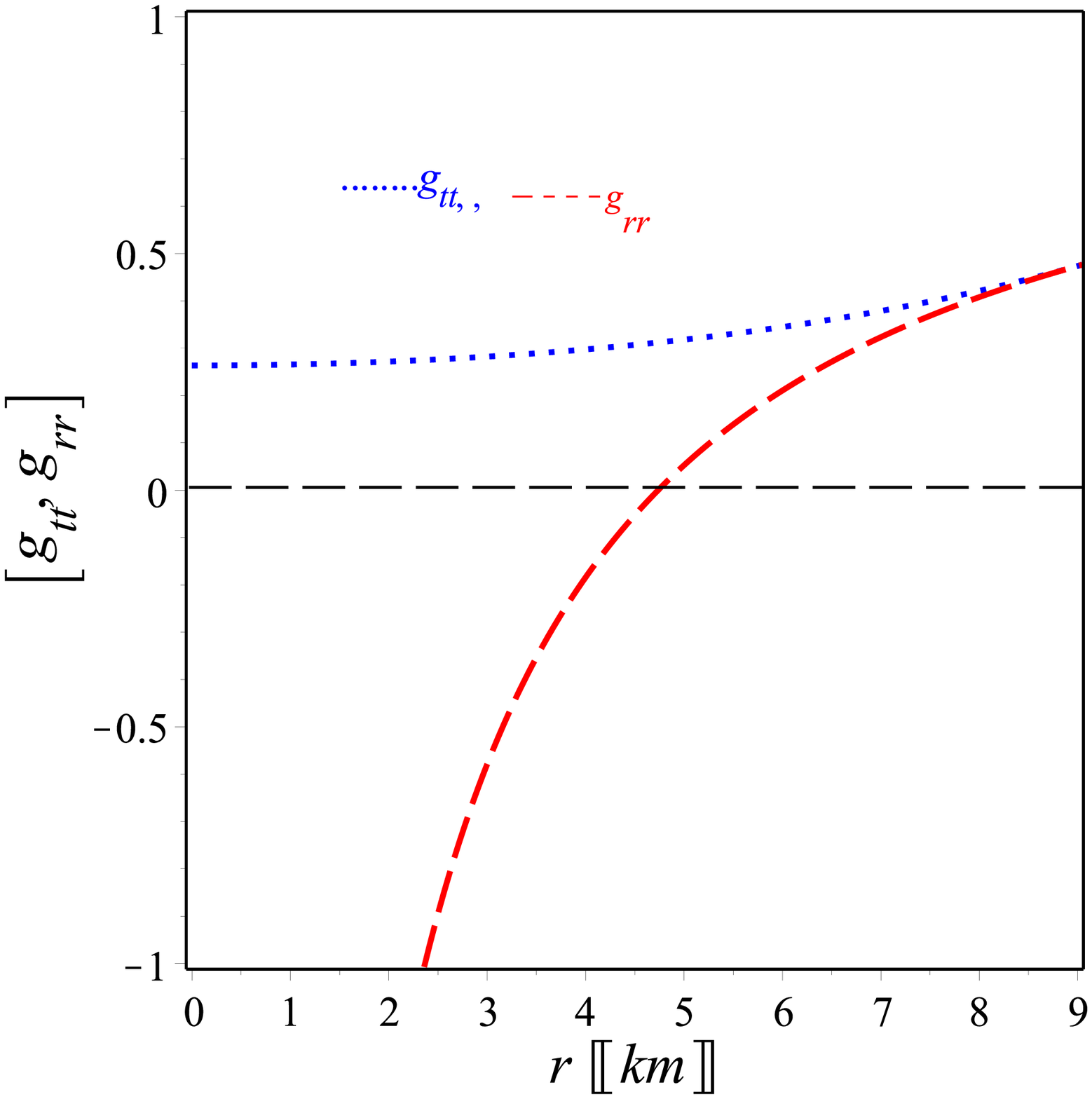}
 \caption {{\it{The temporal component (left graph) and the spatial component
(middle graph) of the metric of the anisotropic solution
(\ref{interiorsolution}), as functions of the radial distance, for various
values of the
parameter $s$, using the  4U 1608-52
mass and radius values. In the right graph we present the smooth matching of
the temporal component with the Schwarzschild
exterior solution.}}}
\label{Fig:4}
\end{figure}

We proceed by investigating the behavior of the metric potentials.
In   Fig.  \ref{Fig:4}    we
present the temporal     and the spatial components, for
various choices of the model parameters. Furthermore, for transparency we
additionally depict  the   smooth matching of
the temporal component with the Schwarzschild
exterior solution. As Fig.  \ref{Fig:4} shows,  the metric
potentials   are both finite
and positive at the center.

\begin{figure}
\centering
 \includegraphics[scale=0.4]{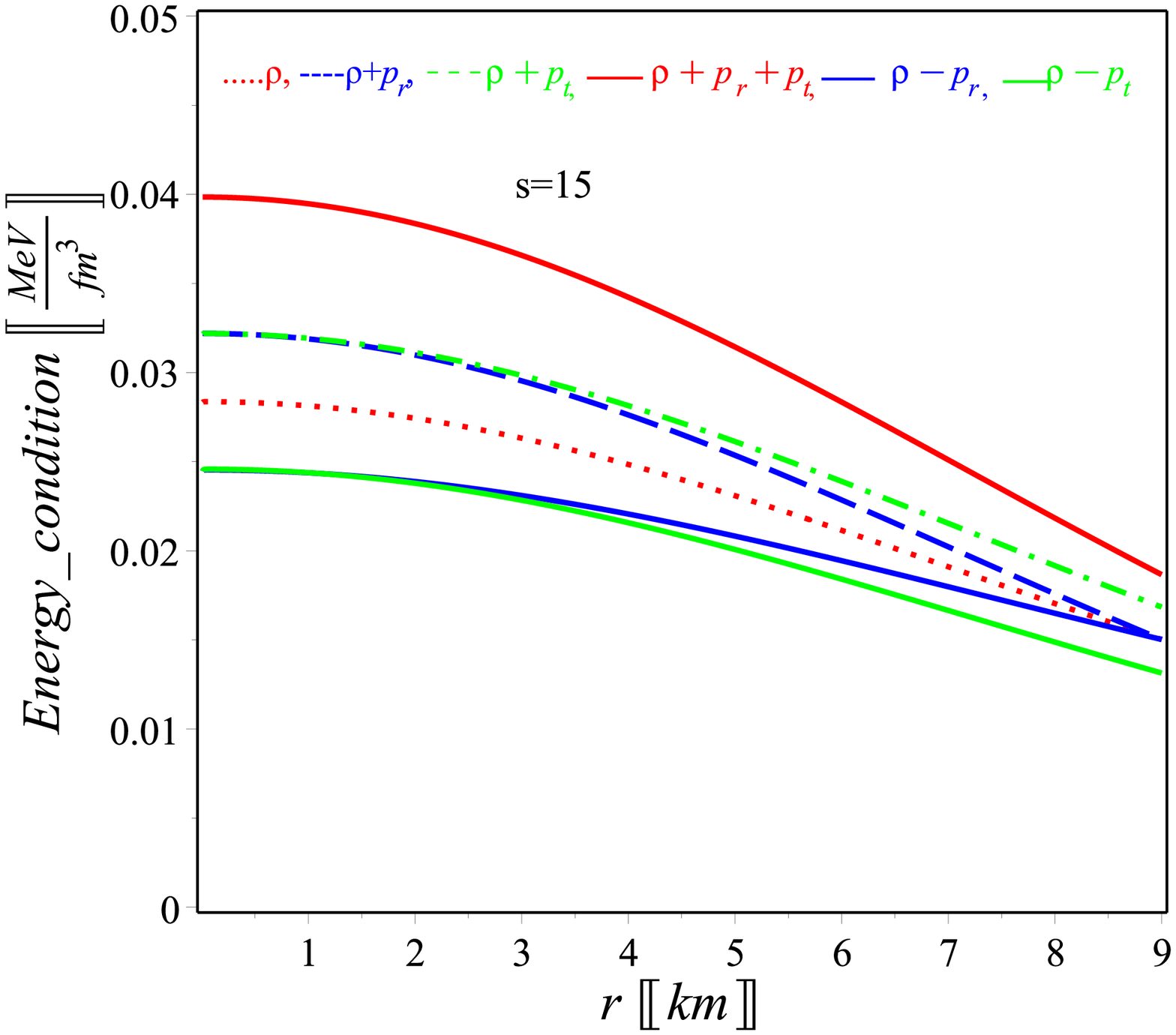}
 \caption {{\it{ The Weak, Null, Strong and Dominant
 energy conditions of the
anisotropic star  solution
(\ref{psol}), as functions of the radial distance, for $s=15$,
  using the  4U 1608-52
mass and radius values.}}}
\label{Fig:5}
\end{figure}

In Fig.  \ref{Fig:5}  we depict the Weak, Null, Strong and Dominant
 energy conditions for $s=15$, showing that they obtain
positive values and thus are all satisfied, as    required for a physically
meaningful stellar model (for other values of $s$ we obtain
similar graphs).

\begin{figure}[ht]
\centering
 \includegraphics[scale=0.26]{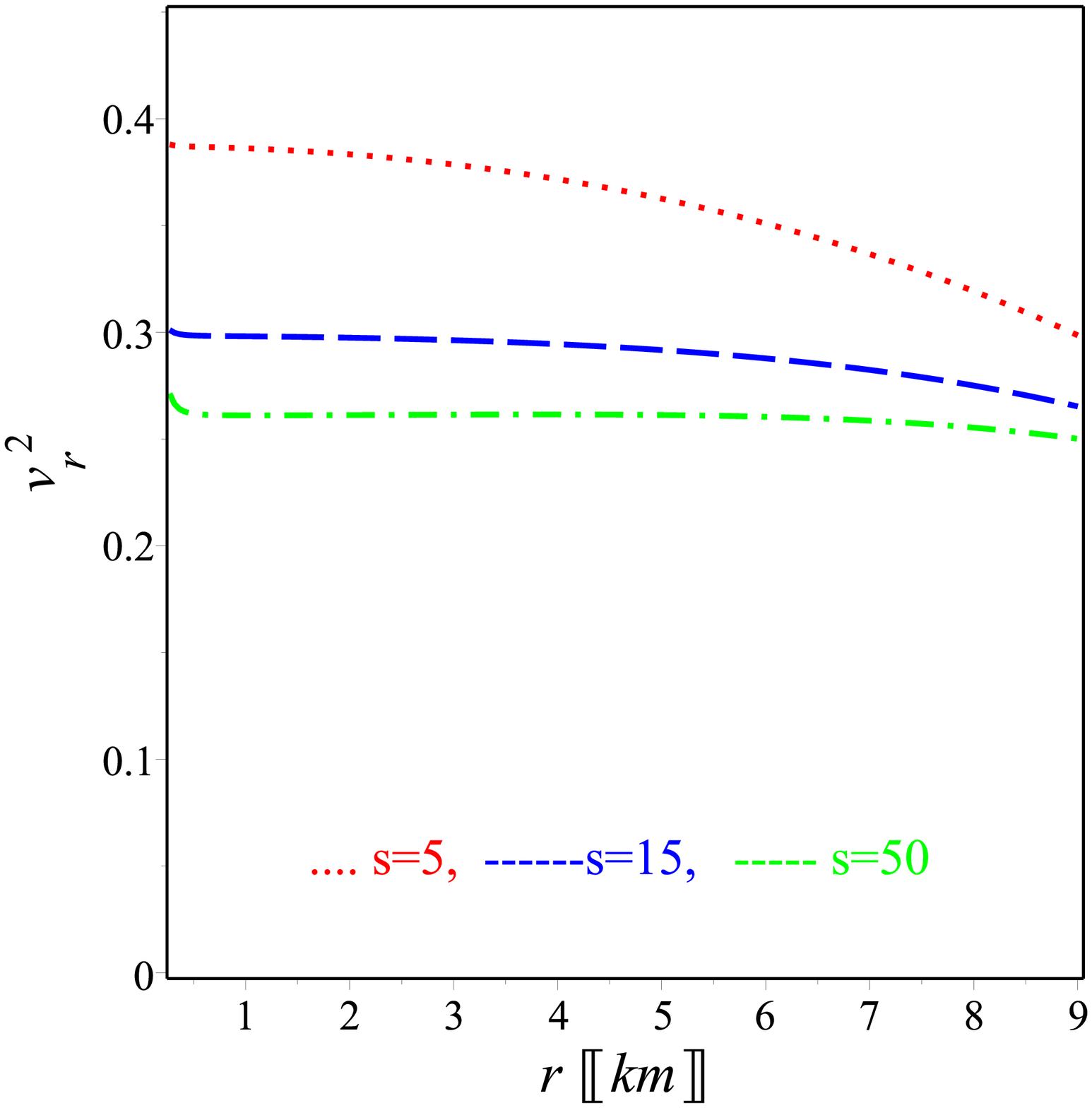}
 \includegraphics[scale=0.26]{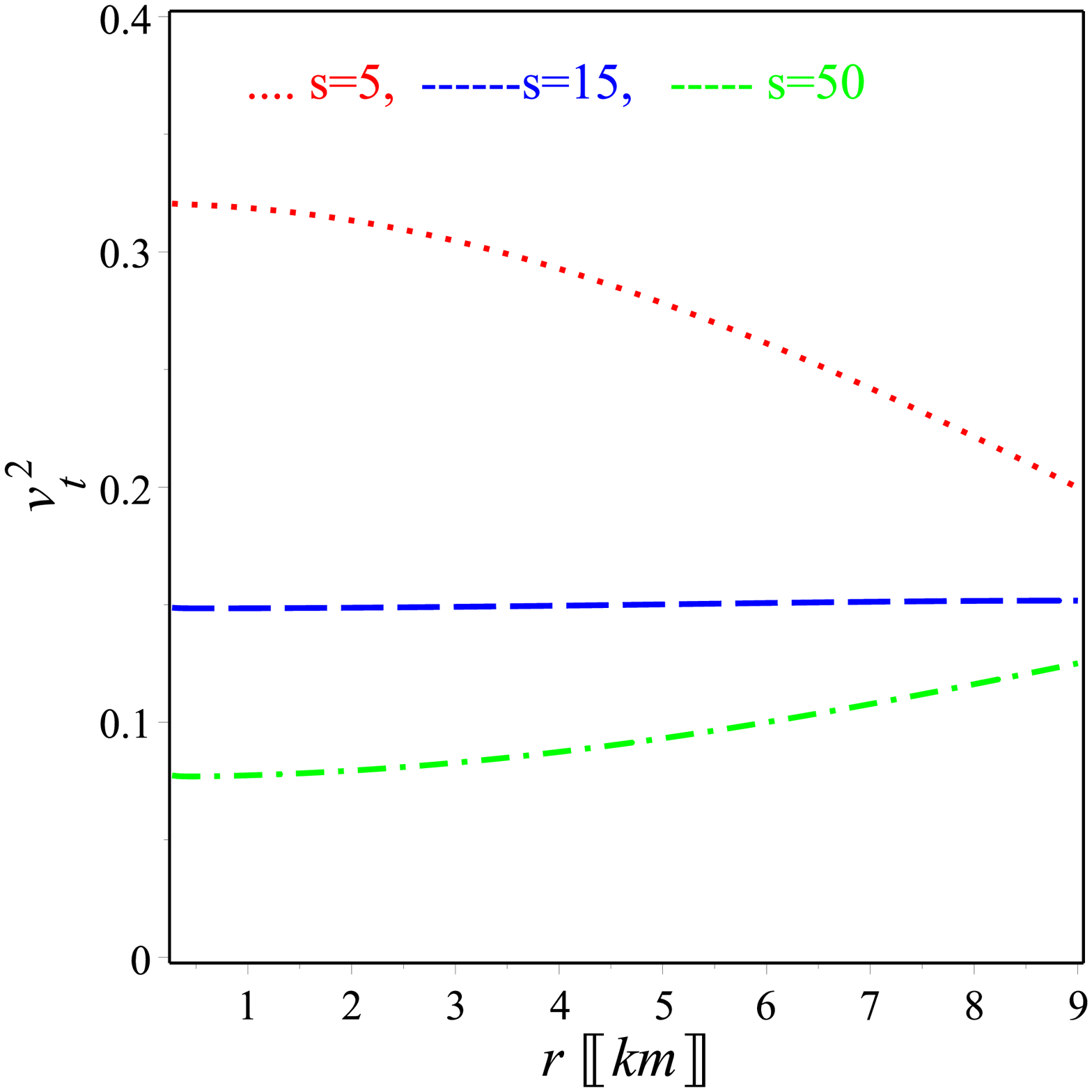}
 \includegraphics[scale=0.26]{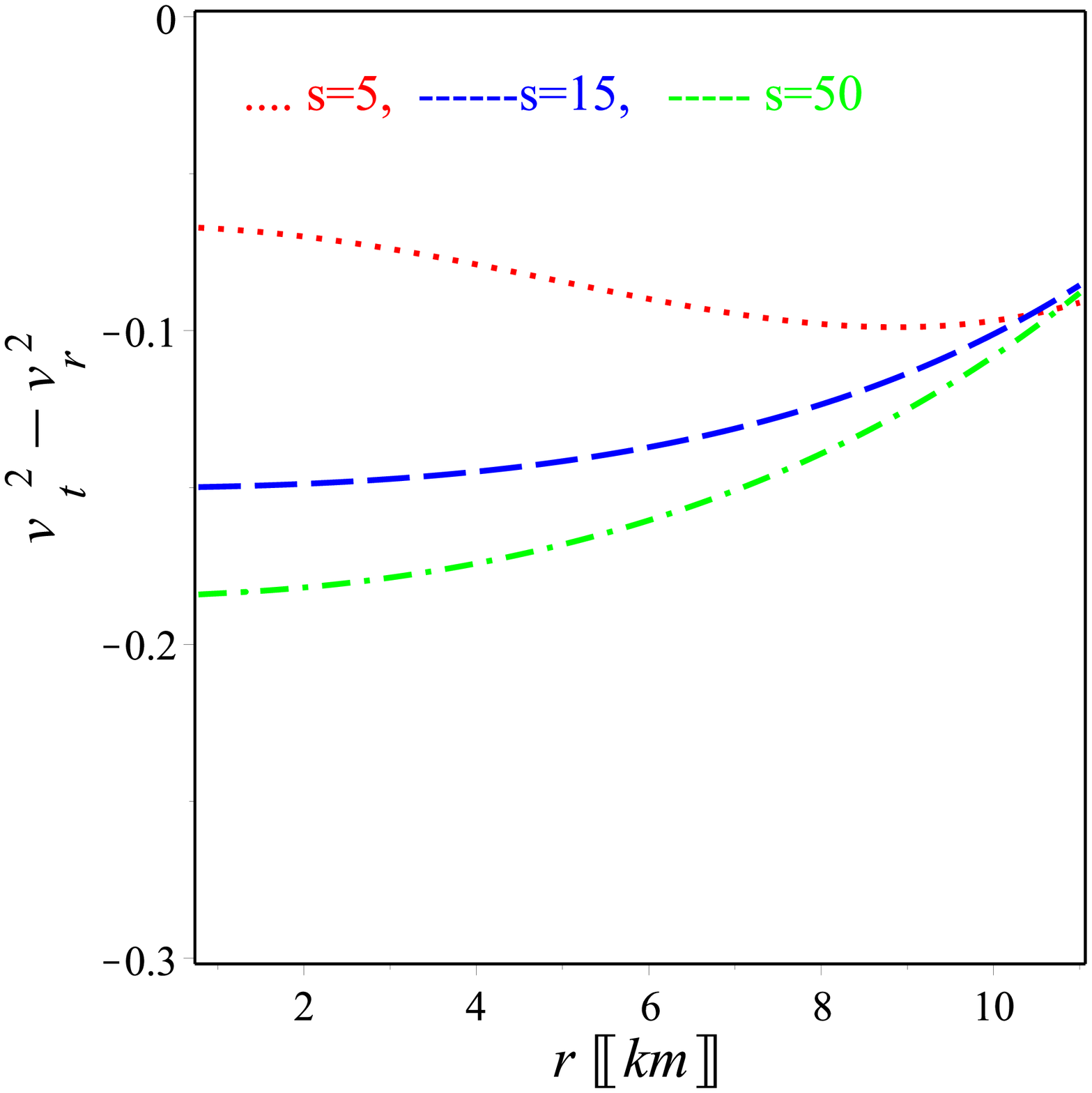}
\caption {\it{{ The radial sound speed square $v_r^2$  (left graph), the
tangentiall sound speed  square $v_t^2$ (middle graph) and the stability factor
$v_t{}^2-v_r{}^2$ (right graph) of the
anisotropic star  solution
(\ref{psol}), as functions of the radial distance, for various values of the
parameter $s$, using the  4U 1608-52    mass and radius values. }}}
\label{Fig:6}
\end{figure}

In the left and middle graphs of Fig. \ref{Fig:6}    we present
 the  radial and tangential sound speed squares, which indeed are
positive and sub-luminal.
Additionally, since   a potentially stable
configuration requires $v_t{}^2-v_r{}^2< 0$
\cite{Herrera:1992lwz,Boehmer:2006ye}, in the right graph of   Fig. \ref{Fig:6}
we depict the stability
 factor  $v_t{}^2-v_r{}^2$, and as
we see it is negative and
hence we conclude that our model is potentially stable
everywhere within the stellar interior for various
values of the parameter $s$.

\begin{figure}[ht]
\centering
\includegraphics[scale=0.26]{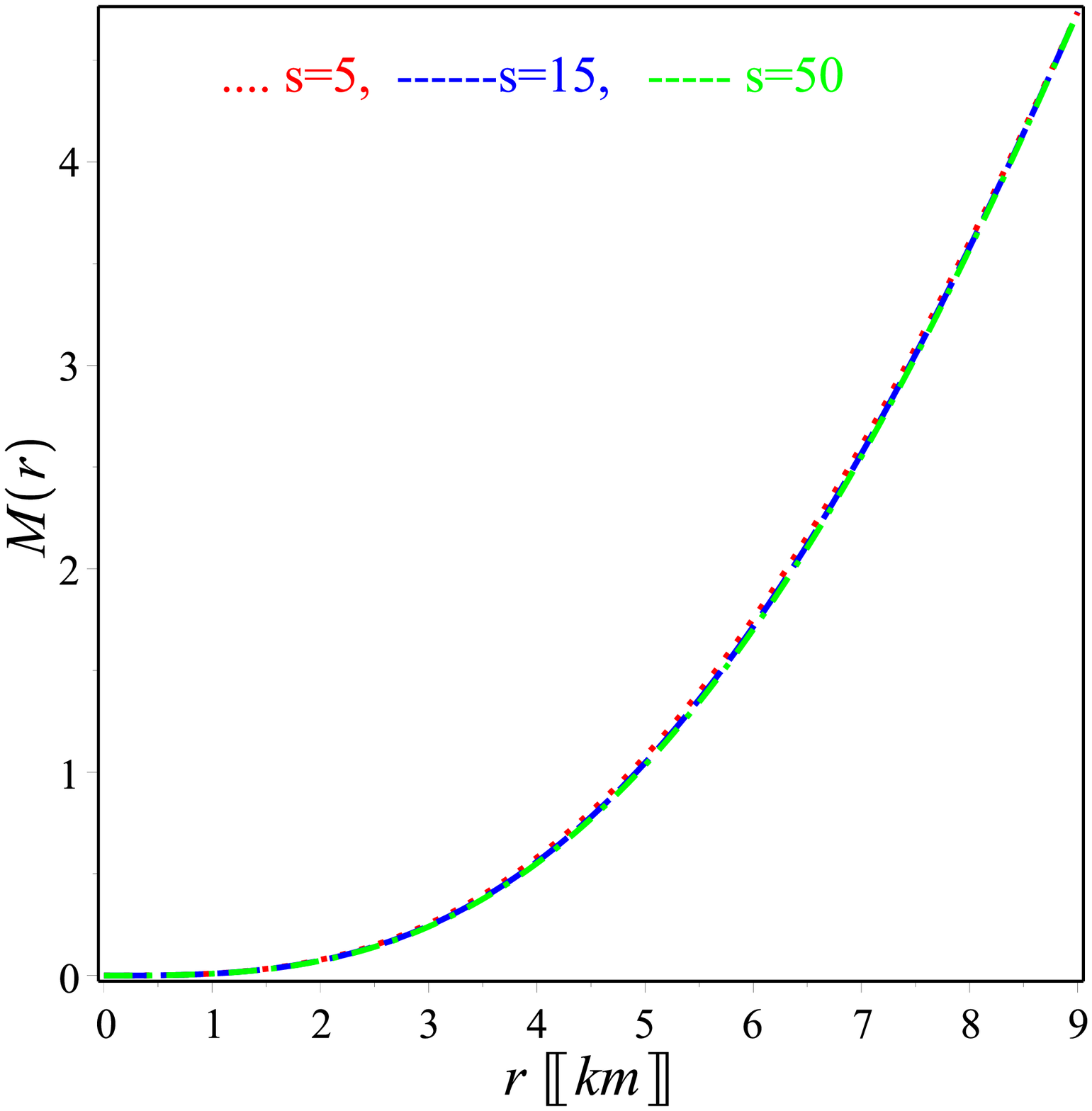}
\includegraphics[scale=0.26]{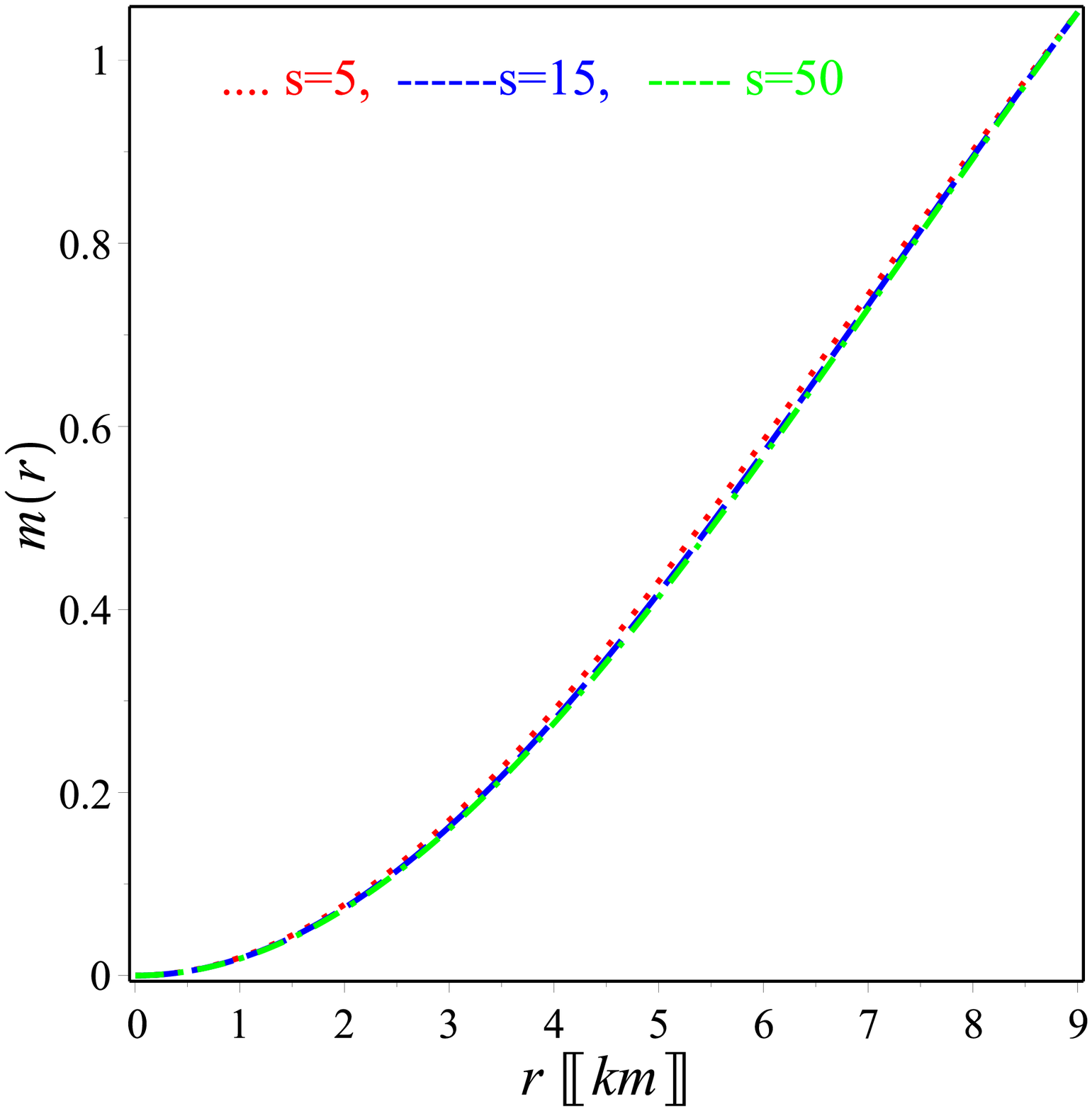}
\includegraphics[scale=0.26]{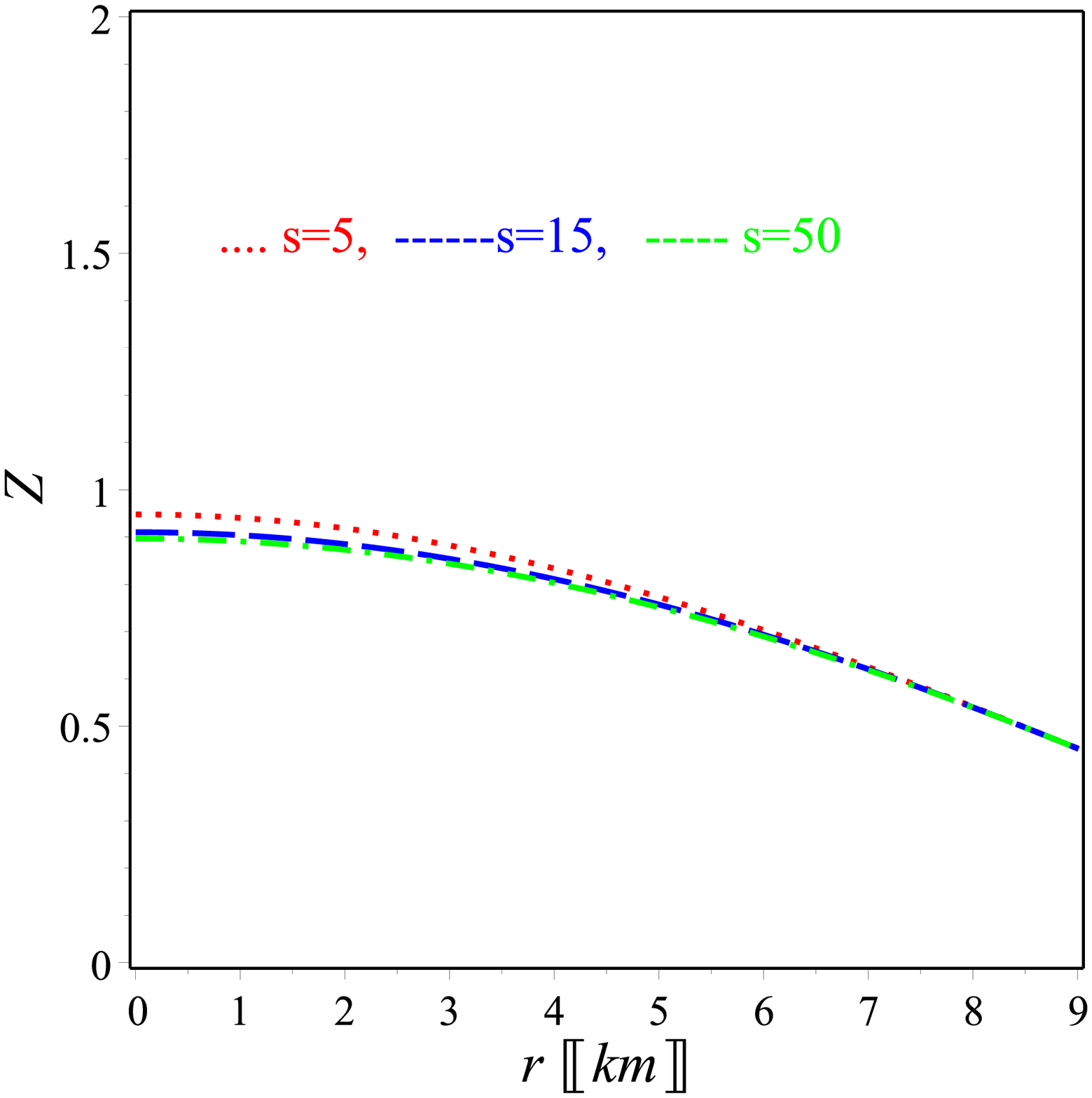}
\caption {\it{{ The mass $M(r)$ of (\ref{mas1}) (left graph), the
compactness $m(r)$ of (\ref{gm1})  (middle graph) and the redshift
(\ref{redshift0}) (right graph) of the
anisotropic star  solution
(\ref{psol}), as functions of the radial distance, for various values of the
parameter $s$, using the  4U 1608-52    mass and radius values. The surface
redshift is $Z_R\approx 0.45$ for all $s$
choices.}}}
\label{Fig:6b}
\end{figure}

The mass function  given  by  (\ref{mas1}) is plotted in the left graph of
Fig. \ref{Fig:6b}, showing that it is a monotonically
increasing function of the radial coordinate and $M(r\rightarrow 0) = 0$.
Furthermore,   the middle graph of
Fig. \ref{Fig:6b}   shows  the behavior of the compactness
parameter (\ref{gm1}), which is increasing. Finally, the radial variation of
the   redshift (\ref{redshift0}) is plotted in the right graph of  Fig.
\ref{Fig:6b}.  We find that the surface  redshift $Z_R\approx 0.45$ for all
$s$
choices, and since the theoretical requirement is $Z_R\leq 2$
\cite{Buchdahl:1959zz}
 we conclude that it is satisfied for    solution
(\ref{psol}).

\section{Stability}\label{stability}

In this section, we are going to discuss the stability issue using two different
techniques; the Tolman-Oppenheimer-Volkoff (TOV) equations and the adiabatic
index. For completeness,   we will also examine the
static case, too.

\subsection{Equilibrium analysis through TOV equation}

\begin{figure}[ht]
\centering
 \includegraphics[scale=0.32]{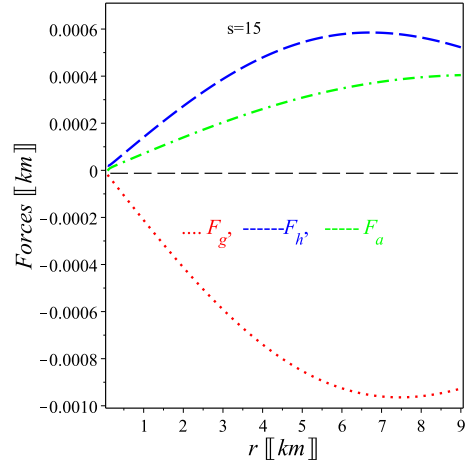}
 \caption {\it{{ The gravitational,  the anisotropic and
the hydrostatic  forces  of the
anisotropic star  solution
(\ref{psol}), as functions of the radial distance,  for
$s=15$, using the  4U 1608-52    mass and radius values.  }}}
\label{Fig:7}
\end{figure}

In this subsection  we are going to discuss the stability of
a general stellar  model. For this goal   we assume hydrostatic equilibrium
through the TOV equation.
Using the TOV equation \cite{Tolman:1939jz,Oppenheimer:1939ne}  as   presented
in \cite{1993GReGr..25.1123P}, we obtain the following  form:
\begin{eqnarray}\label{TOV}
\frac{2[p_t-p_r]}{r}-\frac{M_g(r)[\rho(r)+p_r]\sqrt{h}}{r}-\frac{dp_r}{r}=0\,.
 \end{eqnarray}
Here $M_g(r)$ is  the  mass of the gravitational system at radius $r$, and is
defined through the  Tolman-Whittaker mass formula
\begin{eqnarray}\label{ma}
M_g(r)=4\pi{\int_0}^r\Big({T_t}^t-{T_r}^r-{T_\theta}^\theta-{T_\phi}
^\phi\Big)r^2h_1\sqrt{h}dr=\frac{r(h\,h_1)'}{2h\sqrt{h_1}}\,.
 \end{eqnarray}
Inserting   (\ref{ma}) into (\ref{TOV}) we find
\begin{eqnarray}\label{ma1}
\frac{2(p_t-p_r)}{r}-\frac{dp_r}{dr}-\frac{(h\,h_1)'(\rho+p_r)}{2\sqrt{h\,h_1}}
=F_g+F_a+F_h=0\,,
 \end{eqnarray}
 where $F_g=-\frac{(h\,h_1)'(\rho+p_r)}{2\sqrt{h\,h_1}}$,
$F_a=\frac{2(p_t-p_r)}{r}$, and $F_h=-\frac{dp_r}{dr}$  are the gravitational,
anisotropic and hydrostatic forces,  respectively. The behavior of the TOV
equation for model  (\ref{psol}) is shown in Fig. \ref{Fig:7}, in which the
three different forces are plotted (for other values of $s$ we obtain
similar graphs).
As we observe, the hydrostatic and anisotropic forces are positive, and are
dominated by the gravitational   force which is negative to maintain the
system
in static equilibrium.

\subsection{Adiabatic index}

The stable equilibrium configuration of a spherically symmetric system  can be
studied using the adiabatic index, which is a basic ingredient of the stability
criterion. Considering an adiabatic perturbation, the adiabatic index
$\Gamma$  is defined as
\cite{Chandrasekhar:1964zz,Merafina:2014goa,1989A&A...221....4M,
1993MNRAS.265..533C}:
\begin{eqnarray}\label{a11}
\Gamma=\left(\frac{\rho+p}{p}\right)\left(\frac{dp}{d\rho}\right)\,.
 \end{eqnarray}
 A Newtonian isotropic sphere is in stable
equilibrium if the adiabatic index satisfies $\Gamma>\frac{4}{3}$
  \cite{1975A&A....38...51H}, while for $\Gamma=\frac{4}{3}$ the  isotropic
sphere is  in neutral equilibrium. As it was shown in
\cite{1993MNRAS.265..533C},    for the stability of a relativistic
anisotropic sphere it is required that $\Gamma >\gamma$, where
\begin{eqnarray}\label{ai}
\gamma=\frac{4}{3}-\left[\frac{4(p_r-p_t)}{3\lvert p'_r\lvert}\right]_{max}\,.
 \end{eqnarray}
 Using Eq. (\ref{a11}) and the solution (\ref{psol}), we can find the
expressions for the radial and  tangential adiabatic indices, which are
presented
in   Appendix \ref{AppC}.

In Fig. \ref{Fig:7b}  we draw  $\Gamma_r$ and  $\Gamma_t$ for various values
of $s$. As we can see,
the profile of the radial   and tangential adiabatic indices
are monotonic
increasing functions of $r$ and acquire values greater than $4/3$ everywhere
within the stellar configuration for $s\leq 10$, thus the condition of
stability is satisfied.

\begin{figure}[ht]
\centering
 \includegraphics[scale=0.3]{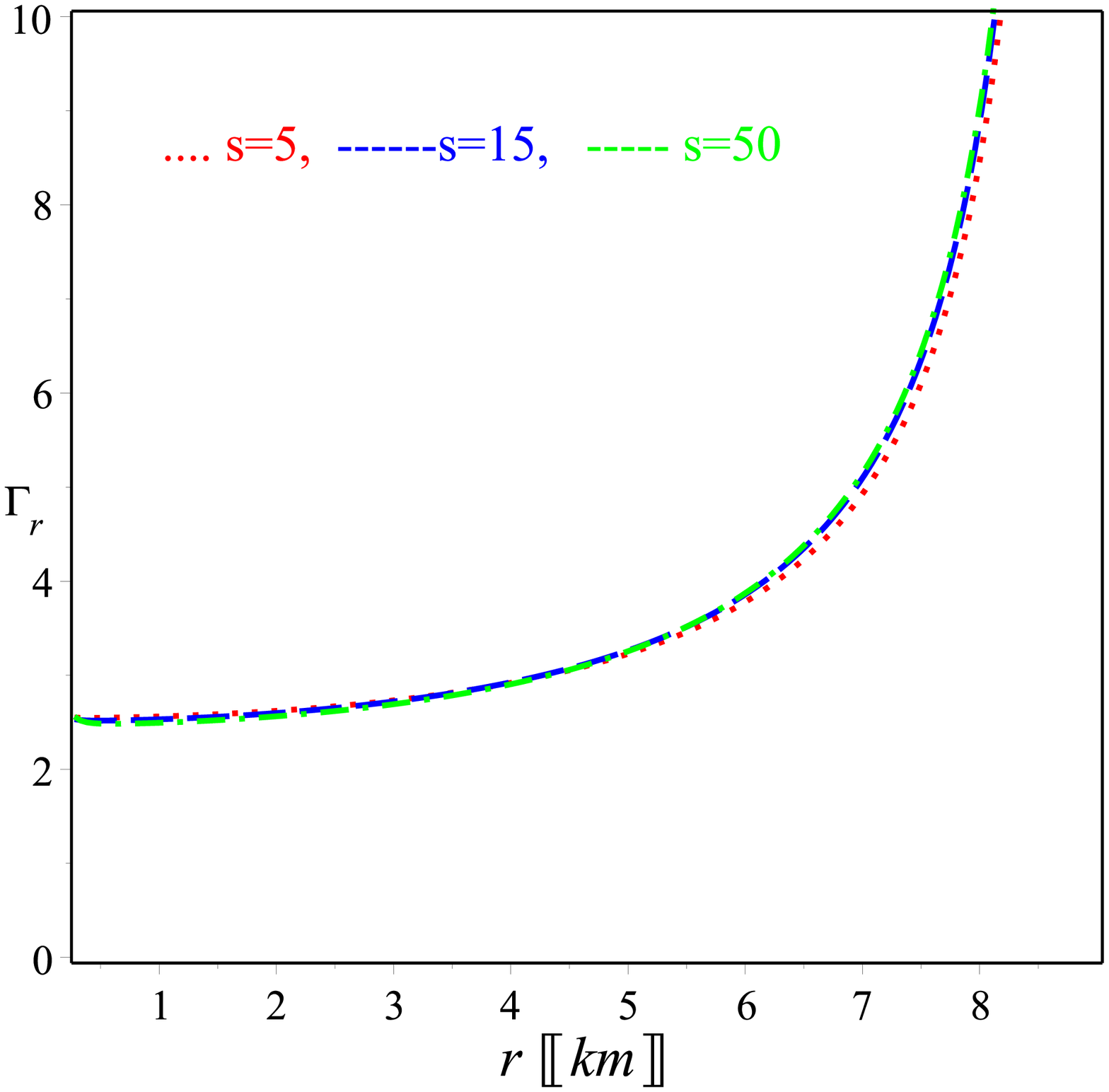}
 \includegraphics[scale=0.3]{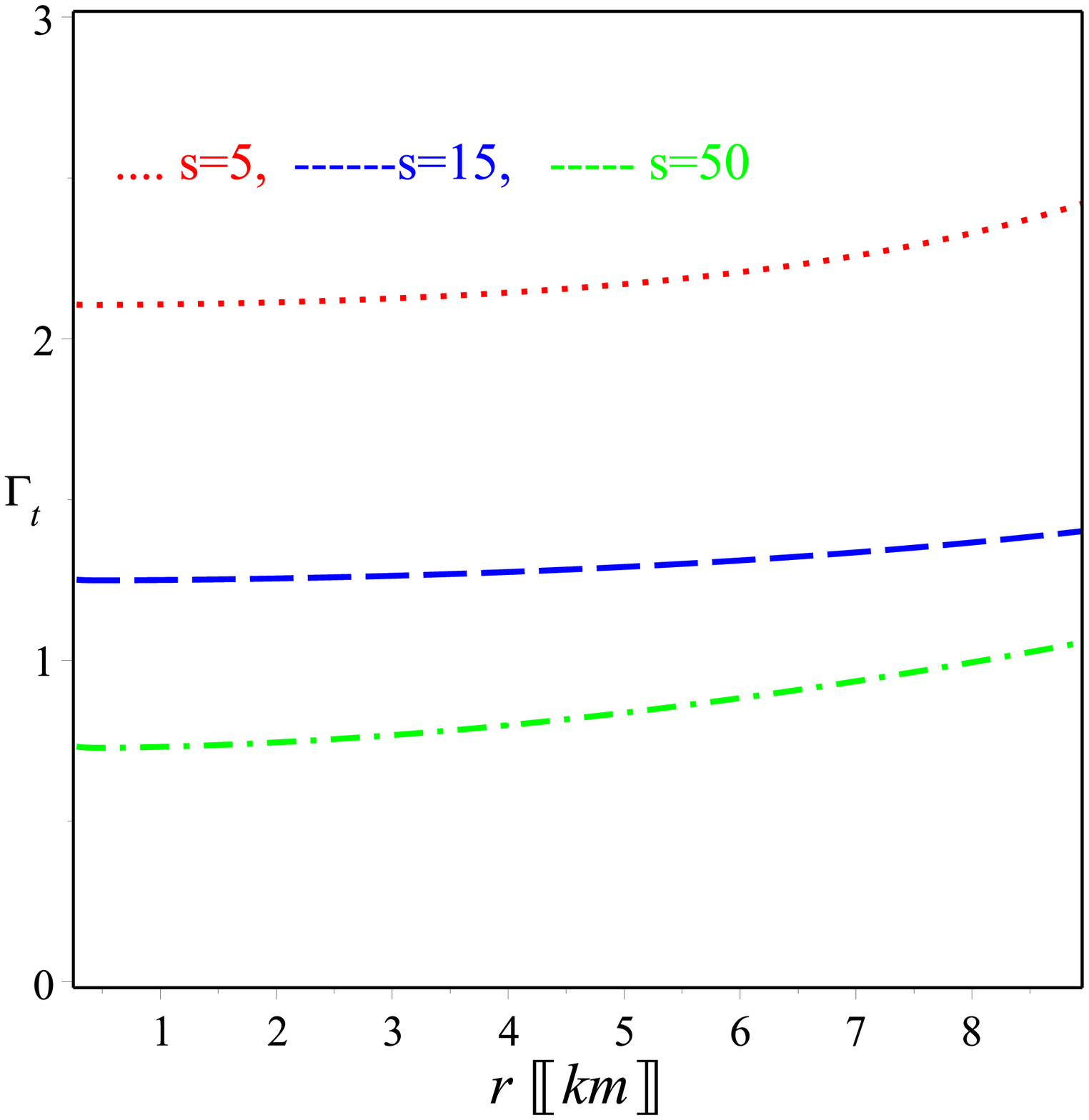}
\caption
{\it{ The radial adiabatic index
(\ref{aic}) (left graph) and the tangential adiabatic  index
(\ref{ait}) (right graph),  versus the radius $r$, for various values of $s$,
using the  4U 1608-52    mass and radius values.}}
\label{Fig:7b}
\end{figure}

\subsection{Stability in the static state}

For completeness, in this subsection we discuss the stability in the static
case. For a stable compact star,  using the mass-central and  mass-radius
expression, as well as the relations for the energy density, Harrison,
Zeldovich, and Novikov  \cite{1965gtgc.book.....H,1971reas.book.....Z} stated
that the gradient of the central density with respect to the mass, should
acquire
positive values, namely $\frac{\partial M}{\partial \rho_{r_0}}> 0$, in order to
have stable configurations.  More precisely, the stable or unstable region
is satisfied for constant mass, namely $\frac{\partial M}{\partial \rho_{r_0}}=
0$ \cite{NewtonSingh:2019bbm}.

\begin{figure}[ht]
\centering
 \label{fig:mrho}
 \includegraphics[scale=0.3]{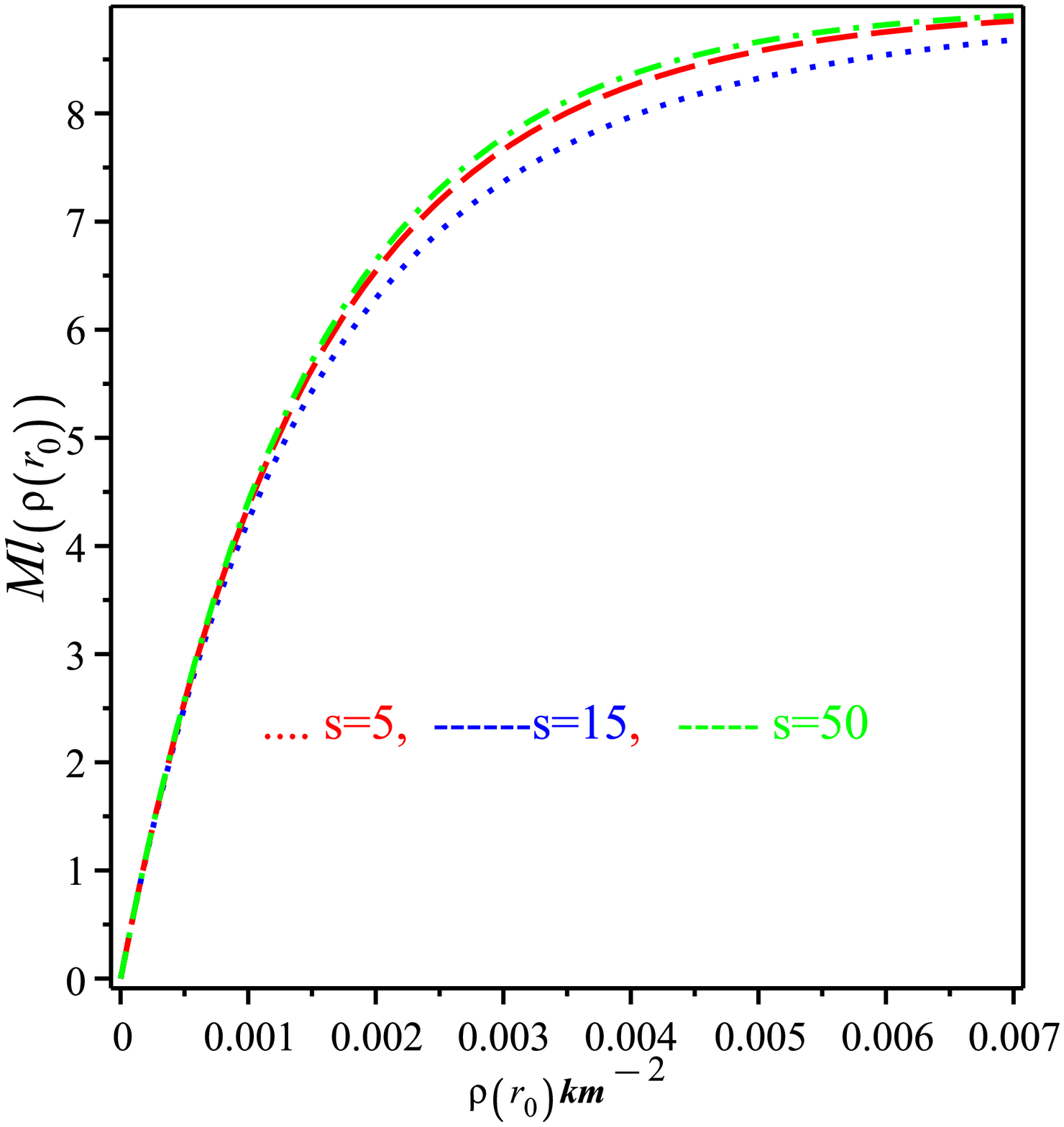}
 \includegraphics[scale=0.308]{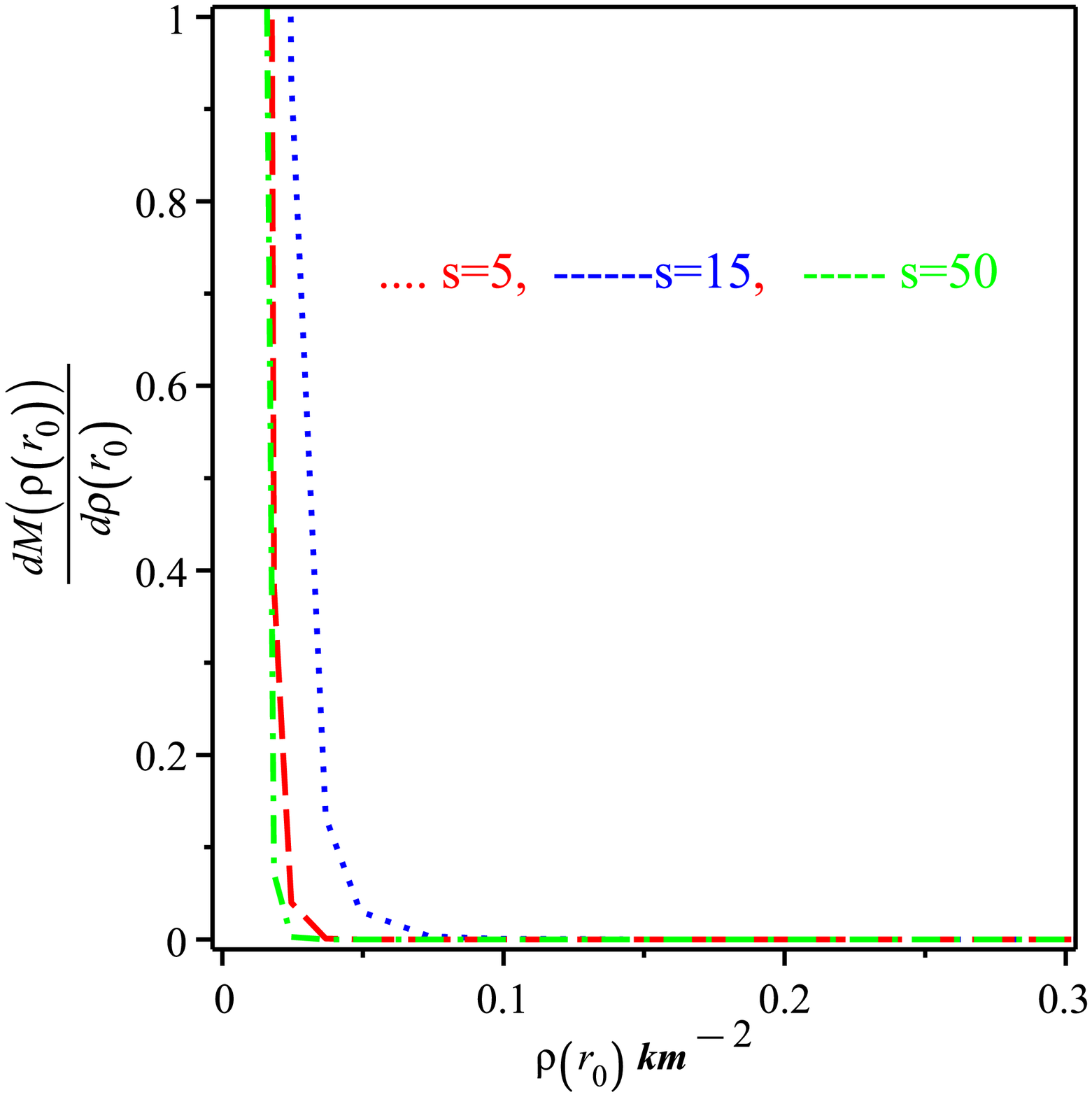}
\caption
{\it{The
mass (\ref{sta}) (left graph) and the  mass gradient  (\ref{sta1})  (right
graph),  as  functions of the central
density $\rho_{r_0}$ for various values of $s$, using the  4U 1608-52    mass
and radius values..}}
\label{Fig:8}
\end{figure}

Let us apply this procedure to our solution
(\ref{psol}).   In this case,  the  central density becomes
$\rho_{_{r_0}}=\frac{3s}{\kappa^2 k^2 } $ and thus we find that
\begin{eqnarray} \label{sta}  M(\rho_{_{r_0}})=l\left[1-\left(1+\frac{\kappa^2
l^2\rho_{r_0}}{3s}\right)^{-s}\right]\,,
 \end{eqnarray}
which finally leads to
 \begin{eqnarray} \label{sta1} \frac{\partial M}{\partial
\rho_{r_0}}=\frac{4\pi\,l^3\left(1+\frac{\kappa^2
l^2\rho_{r_0}}{3s}\right)^{-s-1}}{3}\,.
 \end{eqnarray}
 The above expression  implies that the solution  (\ref{psol}) corresponds to  a
stable configuration since   $\frac{\partial M}{\partial \rho_{r_0}}> 0$
\cite{NewtonSingh:2019bbm}. The behaviors of  the mass and its
gradient    are shown in  Fig. \ref{Fig:8}, and as we observe   the mass
decreases while the gradient-of-mass  increases, as the energy density become
smaller.

\section{Discussion and conclusions}\label{S5}

Mimetic gravity is an interesting modification  obtained   from general
relativity through the  isolation of the conformal degree of freedom in a
covariant way, by applying   a re-parametrization of the physical metric  in
terms of a mimetic field  and an auxiliary metric. The present work aimed to
investigate new anisotropic compact solutions     within mimetic gravity,
since such solutions are known to be very interesting laboratories
of gravity in the strong-field regime.

We derived new classes of anisotropic solutions, by applying the
Tolman-Finch-Skea metric, and a specific anisotropy  not directly depending on
it. Thus, the anisotropy  is positive, it vanishes at the center of the star
and
it has no singularities. Additionally, we determined the involved integration
constants by matching smoothly the interior anisotropic solution to the
Schwarzschild exterior one, requiring additionally that the   pressure must
be vanishing at the boundary of the star.

In order to provide a transparent picture, we used the    data from the 4U
1608-52 pulsar  and we investigated numerically the features of the obtained
anisotropic stars. In particular, we studied the profile of the energy density,
as well as the  radial and tangential pressures, and we showed that they are all
positive and decrease towards the center of the star. Furthermore, we
investigated the anisotropy parameter and the anisotropic force, which are both
increasing functions of the radius, which implies that the latter is
repulsive. Additionally, by examining the radial and tangential
equation-of-state parameters, we showed that they are   monotonically
increasing, and   bounded in the interval [0,1],  which implies that the matter
in our star is not exotic.

Concerning the  metric potentials we saw that they have no singularity, either
at the center of the star or at the boundary, and the matching to the
Schwarzschild
exterior is indeed smooth. Moreover, we examined the  weak, null, strong
and dominant  energy conditions, showing that they are all satisfied in the
interior of the star. Additionally, we examined the radial and tangential sound
speed squares, showing that they are positive and sub-luminal, while we found
that the surface redshift satisfies the requirement $Z_R\leq 2$. Finally, we
provided the profiles for the mass and compactness, which are monotonically
increasing functions of the radius.

In order to investigate the stability of the new anisotropic solutions,  we
applied the  Tolman-Oppenheimer-Volkoff (TOV) equation, we performed the
adiabatic index analysis, and we examined the static case, providing the
profiles of the
 gravitational,    anisotropic and   hydrostatic  forces  of the
star, the radial and tangential adiabatic  indices, as well as the mass
and its gradient, showing that in all cases the star is stable.
 Lastly, for completeness in the Appendix we provided the results for other
pulsar data.

We mention that the obtained solutions do not have a trivial profile for the
mimetic field, and therefore they correspond to novel classes, that do not
exist for general relativity. Hence, the rich behavior of the
aforementioned anisotropic solutions serves as an advantage of mimetic gravity.
It would be interesting to investigate the gravitational wave structure of
possible merges of such solutions, and whether this could provide signatures of
mimetic gravity. Such an analysis will be performed in a future project.

\begin{acknowledgments}
ENS   acknowledges participation in the COST Association Action CA18108
``{\it Quantum Gravity Phenomenology in the Multimessenger Approach (QG-MM)}''.
\end{acknowledgments}

\begin{appendix}

 \section{ The    radial and tangential     sound speeds}
 \label{AppA}

 The    radial and tangential     sound speed squares
can be extracted from (\ref{psol}) as
\begin{eqnarray}
\label{dso2}
&&
\!\!\!\!\! \!\!\!\!\! \!\!\!
v_r{}^2=\frac{dp_r}{d\rho}=
\Bigg\{4 ( {k}^{2}\!+\!{r}^{2} )
^{2}{c_1}^{2} \left[5 ( 1\!+\!s )
 -( {k}^{2}\!+\!{r}^{2} ) ^{2+s}k^{-2s} {r}^{4}
 +{k}^{2} ( s
\!+\!2 ) {r}^{2}\!+\!{k}^{4} \right]
 \left[
 ( {k}^{2}\!+\!{r}^{2} )  ( s\!+\!2 )
 \right] ^{-s}  \nonumber\\
&&
\ \ \ \ \ \ \ \ \ \ \ \ \  \
+(s\! -\!2 )c_1\,c_2
\left[2
 ( {k}^2\!+\!r^2 ) ^{2+s}k^{-2s}-
 ( 2+s^2+12s )
{r}^{4}-2{k}
^{2} ( s\!+\!2 ) {r}^{2}-2{k}^{4} \right]\nonumber\\
&& \ \ \ \ \ \ \ \ \ \ \ \ \  \ \  \ \,\
\cdot {2}^{s/2}  (
{k}^{2}\!+\!{r}^{2} )
\left[  ( {k}^{2
}\!+\!{r}^{2})
( s\!+\!2 )  \right] ^{-s/2}
 \nonumber\\
&& \ \ \ \ \ \ \ \ \ \ \ \ \  \
+
( 2\!-\!s) ^{2}{c_2}^{2}
\Bigg[  ( {k}^{2}\!+\!{r}^{2}) ^
{2}\left(
{\frac {{k}^{2}\!+\!{r}^{2}}{{k}^{2}}} \right) ^{s}+
( -2 {s}^{2}\!-\!3s\!-\!1 ) {r}^{4}-{k}^{2}
( s\!+\!2 ) {r}^{2}\!-\!{k
}^{4} \Bigg]
\Bigg\} \nonumber\\
&&\ \ \ \ \ \ \ \ \ \ \ \ \  \
\cdot\Bigg\{\left\{ c_1\,{2}^{1/2\,s}
( {k}^{2}\!
+\!{r}^{2})
\left[  ( {k}^{2}\!+\!{r}^{2})
( s\!+\!2)
\right] ^{-1/2\,s}-c_2(s \!-\!2 )
 \right\} ^2 \nonumber\\
&&\ \ \ \ \ \ \ \ \ \ \ \ \  \ \ \  \ \cdot\left[
( {k}^{2}\!+\!{r}^{2}) ^{2}
\left( {\frac {{k}^{2}+{r}^{2}}{{k}^{2}}} \right) ^{s}+
( -1\!+\!s\!+\!2{s}^{2}) {r}^{4}-{k}^{2}
( s\!+\!2 )
{r}^{2}-{k}^{4}
\right]\Bigg\}^{-1}\,,
\end{eqnarray}
\begin{table}[ht]
\centering
{\begin{tabular}{@{}ccccccccccccc@{}}\hline\hline
  $s$  && $k\ $ (km$^2$)     &&& $(c_1)^{\frac{1}{s-2}}\ $ (km)  &&&$c_2$ \\
\hline
  1&& $  15.09$ &&&
$ 161$ &&& $ 1.207$\\
  $5$ && $ 38.263$ &&& $133  $&&& $ 0.597$
                                      \\
  $15$ && $ 67.619$&&& $ 268$&&&$ 0.717$
\\
  $20$ && $ 78.273$ &&& $ 335$&&&
$ 0.727$ \\
  $30$ && $ 96.104$ &&& $466$ &&&
$ 0.737$ \\
  $40$ && $ 111.11$ &&& $ 633$ &&&
$ 0.741$ \\
  $50$ && $ 124.316$ &&& $ 724$ &&&
$ 0.744$\\
  $60$ && $ 136.249$ &&& $852$ &&&
$ 0.746$ \\
  $70$ && $ 147.218$ &&& $ 980$ &&&
$ 0.747$  \\
  $80$ && $ 157.425$ &&& $1107$ &&&
$ 0.748$  \\
  $90$ && $ 167.009$ &&& $1234$ &&&
$ 0.748$  \\
  $100$ && $ 176.072$ &&& $1361$ &&&
$ 0.749$  \\
\hline\hline
\end{tabular}}
\caption{The values of the constants $k,\,c_1$ and $c_2$ of
(\ref{b2})-(\ref{kb3}), using the compact star
4U
1724-207, whose observed mass and radius are
$(1.81_{-0.37}^{+0.25})M_{\odot}$ and $12.2\pm1.4$ km
respectively \cite{Miller:2019cac}.} \label{tab1}
\end{table}
\begin{eqnarray}
&&
\!\!\!\!\! \!\!\!\!\! \!\!\!\!\! \!\!\!\!  \!\!
\!\!\!\!\!
v_t{}^2=
\frac{dp_t}{d\rho}={r}^{4}
\Bigg\{
( {k}^{2}\!+\!{r}^{2} )
^{2}{c_1}^{2}
\left[
 ( {k}^{2}\!+\!3\,{r}^{2}) {s}^{2}+ ( -5{k}^{2}\!-\!{r}^{2}
 ) s-4{k}^{2}\!-\!4{r}^{2} \right]
  {2}^{s}
  \left[  ( {k}^{2}\!+\!{r}^{2})
( s\!+\!2)  \right] ^{-s}  \nonumber\\
&&  \ \ \ \ \ \ \ \   +
c_1\,c_2\, \left( 2-s
\right)
 ( {k}^{2}\!+\!{r}^{2})
 \left[
 ( {k}^{2}\!+\!5{r}^{2} ) s-12{k}^{2}-4{r}^{2} \right]
 {2}^{1/2s}s
 \left[  ( {k}^{2}\!+\!{r}^{2} )
 ( s\!+\!2)
 \right] ^{-1/2s} \nonumber\\
&&  \ \ \ \ \ \ \ \
- ( s\!-\!2 ) ^{2}{c_2}^{2}
 \left[
 ( {k}^{2}\!-\!{r}^{2} )
 s-{r}^{2}+3{k}^{2} \right] s
 \Bigg\}
 \nonumber\\
 &&  \ \ \  \cdot
 \Bigg\{ ( {k}^{2}\!+\!{r}^{2} ) \left\{  \left[(
{k}^{2}\!+\!{r}^{2}
) ^{2} - {k}^{2}+{r}^{2} \right] ^{2+s} k^{-2s}
- r^2({r}^{2} \!-\!{k}^{2}
 ) s\right\} \nonumber\\
&&  \ \ \  \cdot
 \left\{c_1\,{2}^{s/2} ( {k}^{2}\!+\!{r}^{2} )
 \left[
 ( {k}^{2}\!+\!{r}^{2} )
 ( s\!+\!2 )  \right] ^{-s/2}
 +c_2 ( 2\!-\!s )  \right\} ^2
 \Bigg\}^{-1}\,.
\end{eqnarray}

\begin{table}[ht]
\centering
{\begin{tabular}{@{}ccccccccccccccccccccc@{}} \hline\hline
s&& {$\rho\lvert_{_{_{0}}}$}  &&{$\rho\lvert_{_{_{l}}}$} & &
{$\frac{dp_r}{d\rho}\lvert_{_{_{0}}}$}   &&
{$\frac{dp_r}{d\rho}\lvert_{_{_{l}}}$}
  && {$\frac{dp_t}{d\rho}\lvert_{_{_{0}}}$} &&
{$\frac{dp_t}{d\rho}\lvert_{_{_{l}}}$}&&{$\Gamma_r\lvert_{_{_{0}}}$}
&&{$\Gamma_t\lvert_{_{_{0}}}$} \\
\hline
 &&&&&&&&\\
 $1$ && $ 0.0132$ && $ 0.005$ && $ 0.56$ &&$ 0.33$
&&$ 0.76$ & &$ 0.179$ &&$ 2.5$    &&$ 3.4$\\
 &&&&&&&&\\
 $5$ && $ 0.01$ && $ 0.0057$ && $ 0.413$   &&$ 0.343$
&
&$ 0.346$&&$ 0.245$ &&$ 2.82$&&$ 2.25$
        \\
  &&&&&&&&\\
  $15$ && $ 0.0098$ && $ 0.0059$ && $ 0.247$  &
&$ 0.228$  & &$ 0.097$&&$ 0.105$
&&$ 2.82$&&$ 1.11$                           \\
   &&&&&&&&\\
 $20$ && $ 0.0097$ && $ 0.0059$ && $ 0.235$
&&$ 0.222$
  &&$ 0.074$&&$ 0.094$    &&$ 2.81$&&$ 0.88$
               \\
  &&&&&&&&\\
  $30$ && $ 0.0097$ && $ 0.0059$ && $ 0.233$  &
&$ 0.217$  & &$ 0.049$&&$ 0.083$
&&$ 2.8$&&$ 0.62$                           \\
   &&&&&&&&\\
   $40$ && $ 0.0097$ && $ 0.0059$ && $ 0.217$
&&$ 0.214$  & &$ 0.037$&&$ 0.077$
&&$ 2.8$&&$ 0.47$                          \\
    &&&&&&&&\\
   $50$ && $ 0.0097$ && $ 0.0059$ && $ 0.213$
&&$ 0.212$   &&$ 0.029$&&$ 0.073$
&&$ 2.8$&&$ 0.38$                           \\
    &&&&&&&&\\
   $60$ && $ 0.0097$ && $ 0.0059$ && $ 0.211$
&&$ 0.21$   &&$ 0.024$&&$ 0.07$  &&$ 2.8$&&$ 0.3$
                    \\
    &&&&&&&&\\
    $70$ && $ 0.0097$ && $ 0.0059$ && $ 0.21$  &
&$ 0.21$  & &$ 0.02$&&$ 0.069$ &&$ 2.8$&&$ 0.27$
                           \\
     &&&&&&&&\\
     $80$ && $ 0.0097$ && $ 0.0059$ && $ 0.21$
&&$ 0.21$   &&$ 0.017$&&$ 0.068$
&&$ 2.8$&&$ 0.2$                            \\
      &&&&&&&&\\
     $90$ && $ 0.0097$ && $ 0.0059$ && $ 0.21$  &
&$ 0.21$   &&$ 0.015$&&$ 0.066$
&&$ 2.8$&&$ 0.2$                          \\
      &&&&&&&&\\
      $100$ && $ 0.0097$ && $ 0.0059$ && $ 0.21$
&&$ 0.21$  & &$ 0.014$&&$ 0.065$
&&$ 2.8$&&$ 0.18$                         \\
\hline\hline
\end{tabular} \label{tab3}}
\caption{Numerical values of   physical quantities
of the anisotropic star  solution
(\ref{psol}),  for various values of the
parameter $s$, using the  4U
1724-207 mass and radius values. The energy
density is measured in Me$\!$V$\cdot$fm$^{-3}$, while all other quantifies are
dimensionless in units where $c=1$.   }
\end{table}

 \section{Analysis using 4U 1724-207 and J0030+0451 pulsars}
 \label{AppB}

In addition to 4U 1608-52, a similar analysis can be developed for other
pulsars. In particular, using the pulsar 4U 1724-207, whose   observed mass
and radius   given by
$(1.81_{-0.37}^{+0.25})M_{\odot}$ and $12.2\pm1.4$ km respectively
\cite{Miller:2019cac}, we obtain the model parameters displayed in
Table 1, and using them we obtain the physical quantities summarized
in Table 2.

Additionally,  in
Table 3 we display the corresponding parameters  for the pulsar
J0030+0451,  whose observed mass and radius is given by
$(1.34_{-0.16}^{+0.15})M_{\odot}$ and $12.71_{-1.19}^{+1.14}$ km respectively
\cite{Riley:2019yda}, and using them we obtain the physical quantities
summarized in Table
4.

\begin{table}[ht]
\centering
{\begin{tabular}{@{}ccccccccccccc@{}} \hline\hline
  $s$   && $k\ $ (km$^2$)     &&& $(c_1)^{\frac{1}{s-2}}\ $ (km)  &&&$c_2$ \\
\hline
 1&& $  17.46$ &&& $  43$ &&&
$ 1.787$\\
 $5$ && $ 42.086$ &&& $ 171$&&& $ 0.521$
                                      \\
 $15$ && $ 73.786$&&& $  302$&&&$ 0.767$
\\
 $20$ && $ 85.33$ &&& $373$&&&
$ 0.786$ \\
 $30$ && $ 104.665$ &&& $ 514$ &&&
$ 0.805$ \\
 $40$ && $ 120.948$ &&& $ 654$ &&&
$ 0.814$ \\
 $50$ && $ 135.285$ &&& $ 794$ &&&
$ 0.819$\\
 $60$ && $ 148.242$ &&& $933$ &&&
$ 0.823$ \\
 $70$ && $ 160.154$ &&& $ 1072$ &&&
$ 0.825$  \\
 $80$ && $ 171.239$ &&& $ 1210$ &&&
$ 0.827$  \\
 $90$ && $ 181.65$ &&& $ 1348$ &&&
$ 0.828$  \\
 $100$ && $ 191.495$ &&& $1486$ &&&
$ 0.83$  \\
\hline\hline
\end{tabular} \label{tab2}}
\caption{The values of the constants $k,\,c_1$ and $c_2$ of
(\ref{b2})-(\ref{kb3}), using the compact star
J0030+0451,  whose observed mass and radius are
$(1.34_{-0.16}^{+0.15})M_{\odot}$ and $12.71_{-1.19}^{+1.14}$ km respectively
\cite{Riley:2019yda}.}
\end{table}

\begin{table}[ht]
\centering
{\begin{tabular}{@{}ccccccccccccccccccccc@{}} \hline\hline
s&& {$\rho\lvert_{_{_{0}}}$}  &&{$\rho\lvert_{_{_{l}}}$} & &
{$\frac{dp_r}{d\rho}\lvert_{_{_{0}}}$}   &&
{$\frac{dp_r}{d\rho}\lvert_{_{_{l}}}$}
  && {$\frac{dp_t}{d\rho}\lvert_{_{_{0}}}$} &&
{$\frac{dp_t}{d\rho}\lvert_{_{_{l}}}$}&&{$\Gamma_r\lvert_{_{_{0}}}$}
&&{$\Gamma_t\lvert_{_{_{0}}}$} \\
\hline
 &&&&&&&&\\
 $1$ && $ 0.0098$ && $ 0.006$ && $ 0.475$ &&$ 0.331$
&&$ 0.675$ & &$ 0.26$ &&$ 3.07$    &&$ 4.36$\\
 &&&&&&&&\\
 $5$ && $ 0.0085$ && $ 0.006$ && $ 0.26$   &&$ 0.23$  &
&$ 0.193$&&$ 0.134$ &&$ 3.83$&&$ 2.85$
       \\
  &&&&&&&&\\
  $15$ && $ 0.0083$ && $ 0.006$ && $ 0.19$  & &$ 0.182$
& &$ 0.04$&&$ 0.046$   &&$ 3.96$&&$ 0.837$
            \\
   &&&&&&&&\\
 $20$ && $ 0.0082$ && $ 0.006$ && $ 0.18$   &&$ 0.176$
&&$ 0.018$&&$ 0.032$    &&$ 3.96$&&$ 0.4$
           \\
  &&&&&&&&\\
  $30$ && $ 0.0082$ && $ 0.006$ && $ 0.17$  & &$ 0.169$
& &$ -0.0044$&&$ 0.0178$   &&$ 3.97$&&$ -0.103$
                 \\
   &&&&&&&&\\
   $40$ && $ 0.0082$ && $ 0.006$ && $ 0.164$
&&$ 0.166$  & &$ -0.016$&&$ 0.01$
&&$ 3.97$&&$ -0.386$                          \\
    &&&&&&&&\\
   $50$ && $ 0.0082$ && $ 0.006$ && $ 0.161$
&&$ 0.164$   &&$ -0.023$&&$ 0.0056$
&&$ 3.97$&&$ -0.567$                           \\
    &&&&&&&&\\
   $60$ && $ 0.0082$ && $ 0.006$ && $ 0.159$
&&$ 0.163$   &&$ -0.028$&&$ 0.003$
&&$ 3.97$&&$ -0.7$                     \\
    &&&&&&&&\\
    $70$ &&$ 0.0082$ && $ 0.006$ && $ 0.158$  &
&$ 0.162$  & &$ -0.031$&&$ 0.0002$
&&$ 3.97$&&$ -0.79$                             \\
     &&&&&&&&\\
     $80$ && $ 0.0082$ && $ 0.006$&& $ 0.156$
&&$ 0.16$   &&$ -0.034$&&$ -0.0015$
&&$ 3.97$&&$ -0.856$                            \\
      &&&&&&&&\\
     $90$ && $ 0.0082$ && $ 0.006$ && $ 0.155$  &
&$ 0.16$   &&$ -0.036$&&$ -0.0028$
&&$ 3.97$&&$ -0.9$                          \\
      &&&&&&&&\\
      $100$ && $ 0.0082$ && $ 0.006$&& $ 0.155$
&&$ 0.16$  & &$ -0.0374$&&$ -0.0039$
&&$ 3.96$&&$ -0.96$                         \\
\hline\hline
\end{tabular} \label{tab4}}
\caption{Numerical values of   physical quantities
of the anisotropic star  solution
(\ref{psol}),  for various values of the
parameter $s$, using the J0030+0451    mass and radius values. The energy
density is measured in Me$\!$V$\cdot$fm$^{-3}$, while all other quantifies are
dimensionless in units where $c=1$.  Note that for
increasing   values of $s$, some quantities acquire nonphysical values, such as
{$\frac{dp_t}{d\rho}\lvert_{_{_{0}}}$},
{$\frac{dp_t}{d\rho}\lvert_{_{_{l}}}$} and {$\Gamma_t\lvert_{_{_{0}}}$}.  }
\end{table}

 \section{ The    radial and tangential  adiabatic index}
\label{AppC}

The adiabatic index of a   spherically symmetric system   is defined as
\cite{Chandrasekhar:1964zz,1989A&A...221....4M,1993MNRAS.265..533C}:
\begin{eqnarray}\label{a112}
\Gamma=\left(\frac{\rho+p}{p}\right)\left(\frac{dp}{d\rho}\right)\,,
 \end{eqnarray}
 and can be applied for the radial and  tangential pressure separately.
  Hence, inserting  the solution (\ref{psol})
into (\ref{a11})   we obtain  the radial adiabatic index  as
 \begin{eqnarray}\label{aic}
&&
\!\!\!\!
\Gamma_r=
2{r}^{2} \left\{ c_1\,{2}^{s/2}
( k^{2}\!+\!{r}^{2} )
(s\!+\!2 )
\left[
( k^{2}\!+\!{r}^{2})
( s\!+\!2)
\right]
^{-1/2\,s}-2\,sc_2\,
( s\!-\!2 )  \right\}\nonumber\\
&& \ \ \ \ \ \ \ \cdot \Bigg\{
( k^{2}\!+\!{r}^{2} ) ^{2}
\left( {\frac
{k^{2}\!+\!{r}^{2}}{k^{2}}} \right) ^{s}+ (s\!+\!2{s}^{2} \!-\!1)
{r}^{4}-k^{2}
( s\!+\!2 ) {r} ^{2}-k^{4} \Bigg\}^{-1}\nonumber
  \end{eqnarray}
\begin{eqnarray}&&
 \ \ \ \ \ \ \
\cdot
\Bigg\{  s^s
\left[ (k^{2}\!+\!{r}^{2} ) ^{2+s}
k^{-2s}-
5( s\!+\! 1)
{r}^{4} -k^{2}
 ( s \!+\!2 ) {r}^{2}
 -k^{4} \right]
 {c_1}^{2} ( k^{2}\!+\!{r}^{2}) ^{2}
 \left[  ( k^{2}\!+{\!r}^{2} )
( s\!+\!2 )
\right] ^{-s}\nonumber\\
&& \ \ \ \ \ \ \  \  \ \,\ -
(s\! -\!2 ) {2}^{s/2}
\left[  2( k^{2}\!+\!{r}^{2} )
^{2+s}
k^{-2s}-
({s}^{2}\!+\!12s+2 ) {r}^{4}-2k^4 -2k^{2}
 ( s\!+\!2 )
 {r}^{2} \right]\nonumber\\
 &&\ \ \ \ \ \ \  \  \ \,\  \ \,\
\cdot c_1 ( k^{2}\!+\!{r}^{2} )
c_2 \left[  ( k^{2}\!+\!{r}^{2} )
( s\!+\!2 )  \right] ^
{-s/2}\nonumber\\
&& \ \ \ \ \ \ \   \ \ \,\ +
(s \!-\!2 ) ^{2}{c_2}^{2}
\left[  ( k ^{2}\!+\!{r}^{2} )
^{2+s} k^{-2s}-(2{s}^{2}\!+\!3s\!+\!1 )
{r}^{4}-k^{4} -k^{2} ( s\!+\!2 ) {r}^{2}
 \right]  \Bigg\} \nonumber\\
&& \ \ \ \ \ \ \ \cdot
 \Bigg\{c_1(k^{2}\!+\!{r}^{2} )
 \left[ ( k^{2}\!+\!{r}^{2})^{1+s}
 k^{-2s}\!-\!k^{2}\!-\!5{r}^{2} \right]
 {2}^{s/2} \left[  ( k^{2}\!+\!{r}^{2})  ( s\!+\!2
)  \right] ^{-s/2}
 \nonumber\\
 && \ \ \ \ \ \ \ \ \
-c_2\Bigg[ ( k^{2}\!+\!{r}^{2} )^{1+s}k^{-2s}
+ ( -2s-1) {r}^{2}\!-\!k^{2} \Bigg]
 ( s\!-\!2 )  \Bigg\}
^{-1}\nonumber\\
&& \ \ \ \ \ \ \
\cdot\Bigg\{ c_1\,{2}^{s/2}
( k^{2}\!+\!{r}^{2})   \left[ ( k^{2}\!+\!{r}^{2} ) ( s\!+\! 2)  \right]
^{-1/2\,s}-c_2( s\!-\!2) \Bigg\}^{-2}\,,
 \end{eqnarray}
 and the tangential adiabatic index as
  \begin{eqnarray}\label{ait}
&&\!\!\!\!\!\!\!\!\!\!\!\!\!\!
\Gamma_t={r}^{2}
\Bigg\{ 4{c_1}^{2} \left[ (3{s}^{2}\! -\!4\!-\!s ) {r}^{2}+
(s^2 \!-\!5s\!-\!4) {k}^{2} \right]
( {k}^{2}\!+\!{r}^{2} ) ^{2}
 \left[ ( {k}^{2}\!+\!{r}^{2} )
 ( s\!+\!2 )  \right] ^
{-s}\nonumber\\
 && \ \ \ \ \ -c_2\,{2}^{s/2}s
(s \!-\!2 )
c_1 ( {k}^{2}\!+\!{r}^{2} )
\left[  ( 5s\!-4\!) {r}^{2} +{k}^{2} ( s\!-\!12 )  \right]  \left[
 ( {k}^{2}\!+\!{r}^{2} )
 ( s\!+\!2 )  \right] ^{-s/2}\nonumber\\
&& \ \ \ \ \
 -{c_2}^{2} s \left( s-2 \right) ^{2}
\left[ {k}^{2} ( s\!+\!3 )-
(
s\!+\!1) {r}^{2} \right] \Bigg\}
 \nonumber\\
 &&\cdot
\Bigg\{ {2}^{s/2}c_1
\left[  ( {k}^{2\!}+\!{r}^{2}) ^{2+s} k^{-2s}+
 ( 3-s ) {r}^{4}+{k}^{2}
 ( s\!+\!2 ) {r}^{2}-{k}^{4} \right]
  \nonumber\\
 && \ \ \    \cdot
 ( {k}^{2}\!+\!{r}^{2} )
 \left[
( {k}^{2}\!+\!{r}^{2} )
( s\!+\!2 )
\right] ^{-s/2}
\nonumber
  \end{eqnarray}
\begin{eqnarray}
 &&\ \ \ \,
 -c_2
 ( s\!-\!2 )
 \left[  ( {k}^{2}\!+\!{r}^{2} ) ^{2+s}k^{-2s}+
( s\!-\!1) {r}^{4}+( 3s\!-\!2 ) {k}^{2}{r}^{2}\!-\!{k}^{4}
 \right]  \Bigg\}  \nonumber\\
 && \cdot    \Bigg\{
 \left[( {k}^{2}\!+\!{r}^{2} ) ^{2+s}
 k^{-2s}\!+\! ( s\!-\!1\!+\!2{s}^{2}) {r}^{4}\!-\!{k}^{2}( s\!+\!2)
{r}^{2}\!-\!{k}^{4} \right] \nonumber\\
 && \ \ \ \cdot  ( {k}^{2}\!+\!{r}^{2} )
\left\{ c_1\,{2}^{1/2\,s}
( {k}^{2}\!+\!{r}^{2})  \left[
( {k}^{2}\!+\!{r}^{2})
(s\!+\!2)
 \right] ^{-1/2s}-c_2
 ( s\!-\!2 )  \right\}\nonumber\\
 &&\cdot
 \left\{  \left[ ( 3s\!-\!4 ) {r}^{2}+ ( s\!-\!4 ) {k}
^{2} \right] {2}^{1/2\,s}c_1\, \left( {k}^{2}+{r}^{2} \right)
 \left[
 ( {k}^{2}\!+\!{r}^{2} )
 ( s\!+\!2 )  \right] ^
{-1/2s}\right. \nonumber\\
&&\left. \ \ \ \
+sc_2 ( k\!-\!r )
( k\!+\!r ) ( s\!-\!2 )  \right\}
\Bigg\}^{-1}\,.
\end{eqnarray}

\end{appendix}

\section*{Data Availability Statement}
No Data associated in the manuscript.


%

\end{document}